\documentclass[a4paper,11pt]{article}

\usepackage{amsthm,amsmath}
\usepackage[round]{natbib}
\usepackage[colorlinks,citecolor=blue,urlcolor=blue]{hyperref}
\usepackage{amssymb}
\usepackage{graphicx}
\usepackage{amsfonts}
\usepackage{amsbsy}
\usepackage{mathrsfs}
\usepackage{lscape}
\usepackage{float}
\usepackage{multirow}
\usepackage{tablefootnote}
\usepackage{threeparttable}
\usepackage{pdflscape}
\usepackage{enumerate}
\usepackage{pdfpages}
\usepackage{authblk}
\usepackage{dblfloatfix}
\usepackage{graphicx}
\usepackage[caption=false]{subfig}
\usepackage{adjustbox}
\usepackage{booktabs}
\RequirePackage{fix-cm}
\usepackage{enumerate}
\usepackage{pdfpages}
\usepackage{comment}
\usepackage{tabularx,ragged2e,booktabs,caption}
\usepackage{lipsum}
\usepackage{mathtools}
\usepackage{cuted}

\usepackage{algorithmicx}
\usepackage[ruled]{algorithm}
\usepackage{algpseudocode}
\usepackage{color}
\usepackage{array}
\usepackage{dblfloatfix}
\newcolumntype{H}{>{\setbox0=\hbox\bgroup}c<{\egroup}@{}}
\usepackage{amsmath}
\usepackage{bm}
\usepackage[misc,geometry]{ifsym}

\theoremstyle{plain}

\theoremstyle{definition}

\graphicspath{{.}{./Figure/}}

\newcommand{\bbeta}{\boldsymbol{\beta}}

\newcommand{\bX}{\mathbf{X}}

\newcommand{\bld}{{\boldsymbol{\Lambda}}}

\begin{document}
	
	
	\title{Quantifying effect of geological factor on distribution of earthquake {occurrences} by inhomogeneous Cox processes}
	\author[1]{Achmad Choiruddin}
	\author[1]{Aisah}
	\author[1]{Finola Trisnisa}
	\author[1]{Nur Iriawan}
	\affil[1]{Department of Statistics, Institut Teknologi Sepuluh Nopember, Indonesia \\ choiruddin@its.ac.id}
	\maketitle
	
	\begin{abstract}
		Research on the earthquake occurrences using a statistical methodology based on point processes has mainly focused on the data featuring earthquake catalogs which ignores the effect of environmental variables. In this paper, we introduce inhomogeneous versions of the Cox process models which are able to quantify the effect of geological factors such as subduction zone, fault, and volcano on the major earthquake distribution in Sulawesi and Maluku, Indonesia. In particular, we compare Thomas, Cauchy, variance-Gamma, and log-Gaussian Cox models and consider parametric intensity and pair correlation functions to enhance model interpretability. We perform model selection using the Akaike information criterion (AIC) and envelopes test. We conclude that the nearest distances to the subduction zone and volcano give a significant impact on the risk of earthquake occurrence in Sulawesi and Maluku. Furthermore, the Cauchy and variance-Gamma cluster models fit well the major earthquake distribution in Sulawesi and Maluku.

	\end{abstract}

	\section{Introduction}
	\label{intro}
	Statistical methodology based on point processes has become one of the standard tools for the modeling of earthquake occurrences, e.g. \citet{vere1970stochastic,ogata1988statistical,ogata1999seismicity,matsu2005point,siino2018joint} and references therein. To analyze such data, the pattern of earthquake locations (resp. location and time occurrences) is regarded as a realization of spatial (resp. spatio-temporal)  point process. Furthermore, marked point processes can be employed if e.g. the magnitude or depth of each earthquake epicenter is included in the analysis. We refer these features (coordinate, time occurrence, magnitude, depth) to as earthquake catalogs.
	
	Within the point process framework, earthquake da-ta analysis is usually conducted using two types of modeling, namely intensity and conditional intensity-based modeling. Conditional intensity-based model might belong to Gibbs or Hawkes point processes \citep[e.g.][]{ogata1999seismicity,anwar2012implementation} while intensity-based modeling includes Poisson and Cox point processes \citep[e.g.][]{turkyilmaz2013comparing,siino2018joint,aisah2020onthe}. The model using conditional intensity or intensity is complementary and each has its virtues, see e.g. review by \cite{moller2007modern}.  For example,  Gibbs models have clear interpretations in terms of their Papangelou conditional intensities, while their intensity functions are not tractable. In contrast, Cox processes have tractable intensity and second-order intensity functions which are advantageous in terms of marginal interpretations.
	
	{Traditionally, using either intensity or conditional intensity, the analysis is focused only on the seismic catalogs data which often assumes stationary models \citep{ogata1988statistical,zhuang2002stochastic,turkyilmaz2013comparing}. However, with the advancement of technology and huge investment in data collection, more data information is easily accessed and could be included in the analysis to enhance model interpretability and improve model prediction. For example, we could involve geological factors to study their relation with and predict the distribution of earthquake epicenters. To take the effect of geological variables into account, the traditional modeling does not work so the modified intensity or conditional intensity models should be considered. Recently, this problem has been studied by \cite{anwar2012implementation,siino2017spatial} by considering the distance to the nearest subduction zones, faults, or volcanoes as geological variables. However, such studies only focus on the conditional intensity-type modeling, especially the ones belonging to the class of inhomogeneous Gibbs point processes such as Strauss and Geyer saturation processes.}
	
	In this paper, we extend the applicability of point processes to model {earthquake occurrences} along with geological factors by considering intensity-based modeling which belongs to Cox processes. More precisely, we employ the inhomogeneous versions of the variants of Neyman-Scott Cox processes (NSCP) such as Thomas, Cauchy, and variance-Gamma processes and the log-Gaussian Cox processes (LGCP) {to model earthquake locations with magnitudes of 5 or greater} in Sulawesi and Maluku, Indonesia. 
	Cox models have been considered and are standard for earthquake modeling, especially the Thomas process and LGCP  \citep[e.g.][]{vere1970stochastic,turkyilmaz2013comparing,siino2018joint}. {Nevertheless, no study considering Cox models include geological variables in the analysis, thus extending these models to be able to capture the effect of geological factors is a major importance.} Furthermore, to our knowledge, the Cauchy and variance-Gamma models are new proposals for earthquake modeling (see Sections~\ref{sec:NSCP}-\ref{sec:LGCP}).
	
	One particular challenge with Cox processes is the intractable likelihood function, so the estimation using e.g. the Markov Chain Monte Carlo is computationally expensive \citep{moller2003statistical}. {Because of that, we do not evaluate directly the likelihood of the corresponding Cox processes, but we instead follow \citet{guan2006composite} and \citet{tanaka2008parameter} by constructing and maximizing the second-order composite likelihood  or palm likelihood. These techniques are computationally cheap and guarantee the statistical properties of the estimator (consistent, asymptotically normal).} Therefore, this could add a toolbox for earthquake analysis with external variable and users are free to choose between conditional intensity-based and intensity-based modeling.
	
	The rest of the article is organized as follows. We detail in Section~\ref{sec:data} the study area and data description. We present in Section~\ref{sec:method} the statistical methodology and discuss our results in Section~\ref{sec:result}. Concluding remarks are given by Section~\ref{sec:conl}.

	\section{Study area and data description}
	\label{sec:data}
	\begin{figure*}[htbp!]
		\centering
		\begin{tabular}{c}
			\includegraphics[width=0.87\textwidth]{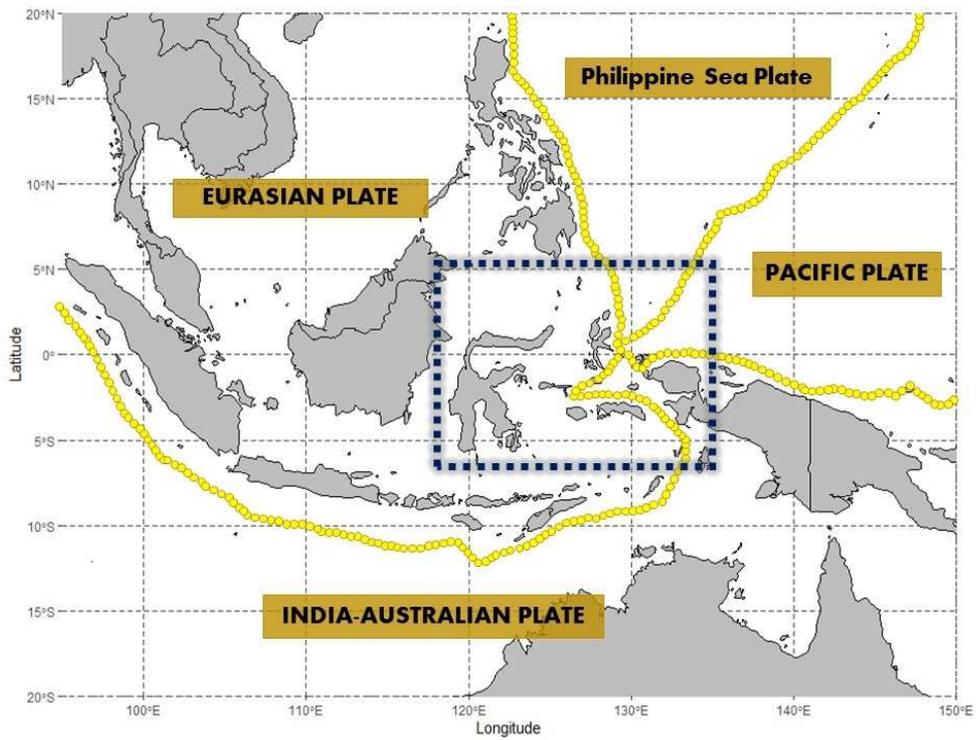}\\
			\includegraphics[width=.94\textwidth]{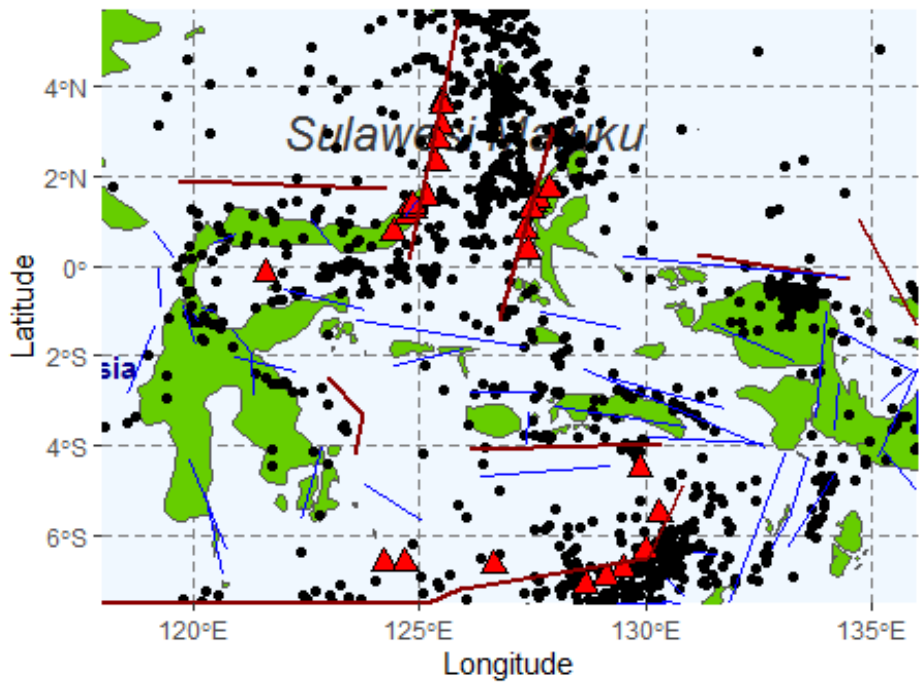}
		\end{tabular}
		\caption{Top: Map of active tectonics in Indonesia. Bottom: Locations of Sulawesi-Maluku earthquake occurrences (black dot) with M$\geq5$ during 2009 – 2018 along with locations of subduction zone (dark red line), fault (blue line), and volcano (red triangle). }
		\label{fig:plate}      
	\end{figure*}
	
	The Indonesian archipelago is located at the junction of the Eurasian, India-Australian, Pacific, and Philippine Sea plates (Figure~\ref{fig:plate} top). The first three plates are major and are constantly moving and colliding. The India-Australian plate moves relatively northward and infiltrates into the Eurasian plate while the Pacific plate moves to the west and collides the India-Australian plate \citep{hamilton1979tectonics,bock2003crustal}. This causes Indonesia to be an earthquake-prone area.

	\renewcommand{\arraystretch}{1.25}
	\begin{table*}[htbp!]
		\caption{List of the geological covariates considered in this study}
		\centering
		\begin{tabular}{clll }
			\hline
			Variable & Description & Unit & Reference\\
			\hline
			$z_1$ & Distance to the nearest subduction & $100$ km & \citep{bilek2018subduction}\\
			$z_2$ & Distance to the nearest volcano & $100$ km & \citep{eggert2009volcanic} \\
			$z_3$ & Distance to the nearest fault & $100$ km &  \citep{luo2019slow} \\
			\hline
			\label{table:covs}
		\end{tabular}
	\end{table*}
	
	Sulawesi and Maluku are part of Indonesia marked with a dashed line in blue located at $D = [-8.33, 6.33] \times [130.98,150.96]$ (100 km)$^2$  (Figure~\ref{fig:plate} top). This region is the only meeting zone of more than two plates in Indonesia, known as the triple junction,  which leads to a high risk of earthquakes. \cite{BMKG2019} records that almost a half of the total earthquakes appearing in Indonesia during 2009-2018 occurs in Sulawesi and Maluku.
	
	We focus in this study on the earthquake occurrences in Sulawesi and Maluku where $m=1080$ events are observed with {magnitudes $\geq 5$} during 2009-2018. We include three geological factors to study their relation to the earthquake distribution, i.e., the subduction zone, fault,  and volcano. {\citet{anwar2012implementation,siino2017spatial} have considered these three factors for earthquake modeling using Gibbs processes and found that the relation between these factors and earthquake events exist. In addition, these geological variables could trigger seismic activities which lead to earthquake occurrences (see Table~\ref{table:covs} and paragraph below)}.
	
	Figure~\ref{fig:plate} (bottom) depicts the distribution of earthquake locations in Sulawesi-Maluku along with subduction zones, faults, and volcanoes. The Sulawesi-Maluku earthquakes tend to occur at the area close to the subduction zones, faults,  and volcanoes. Subduction is a boundary of two colliding plates where one of them is infiltrated into the bowels of the earth and the other is raised to the surface. Major earthquakes often occur near the subduction zones \cite[e.g.][]{bilek2018subduction}. Major subduction zones are observed in the north and south of Sulawesi and Maluku, in particular the North Sulawesi, West and East Molucca subductions in the north and Wetar Back Arc in the south. In these subduction zones volcanoes also exist, in particular are the Sangihe, Halmahera, and Timor. Note that active volcanoes could also trigger earthquakes \citep{eggert2009volcanic}. In addition to the subduction zone and volcano, earthquakes often occur at the area close to a fault \citep{luo2019slow, liu2019spatial}. A fault is a boundary of two fractions of moving earth's crust which occurs typically at the relatively weak or rift area. At least 48 faults are observed in the Sulawesi and Maluku \citep{PSGN2017}, which in particular the Palu-Koro and Matano faults movement causes major earthquakes in Central Sulawesi \citep{baath1979some}. In addition to the spatial trend due to geological variables, the distribution of earthquake locations tend to be clustered possibly due to mainshock and aftershock activities. We investigate in this paper the inhomogeneity due to geological factors and clustering effect using various Cox processes models.

	\section{Methodology} \label{sec:method}
	Let $\bX$ be a spatial point process on $\mathbb{R}^2$. For a bounded observation domain $D \subset \mathbb{R}^2$, $|D|$ denotes the area of observation domain. A realization of $\bX$ in $D$ is thus a set $\mathbf{x}=\{x_1, \ldots, x_m\}$, where $x \in D$ and $m$ is the observed number of points in $D$. In our study, the set $\mathbf{x}$ represents the collection of earthquake coordinates in a region $D$ and $m$ represents the number of earthquake occurrences. If the intensity $\lambda$ and second-order product density $\lambda^{(2)}$ of a point process $\bX$ exist, the pair correlation function $g$ is defined by
	\begin{align} \label{eq:paircorr}
		g(u,v)=\frac{\lambda^{(2)}(u,v)}{\lambda(u)\lambda(v)}, \quad u,v \in D.
	\end{align}
	The pair correlation function is a measure of the departure of the model from the Poisson point process (detailed in the next paragraph) for which $g=1$. An alternative summary function to measure the spatial interaction among earthquake occurrences is the $K$-function, which could be defined as
	\begin{align} \label{eq:K}
		K(r)=2\pi \int_0^r s g(s) \mathrm{d}s, \quad r=\|u-v\|.
	\end{align}
	Further details on spatial point processes could be found in \citet{moller2003statistical,baddeley2015spatial}.
	
	A point process $\mathbf{X}$ on $D$ is a Poisson point process with intensity function $\lambda$ if:
	\begin{enumerate}
		\item for any bounded $B \subseteq D$, the number of points in $B, N(B),$ follows Poisson distribution,
		\item conditionally on $N(B)$, the points in $\mathbf{X} \cap B$ are i.i.d. with joint density proportional to $\lambda(u)$, $u \in B$.
	\end{enumerate}
	For Poisson point processes, $\lambda^{(2)}(u,v)=\lambda(u)\lambda(v)$, so $g(u,v)=1, \forall u,v \in D$.
	
	We consider two classes of Cox processes to model earthquake occurrences in Sulawesi and Maluku. A Cox process is an extension of a Poisson point process, obtained by considering the intensity of the Poisson point process as a realization of a non-negative random field. If the conditional distribution of $\bX$ given $\bld=\{\Lambda(u): u \in D\}$ {(the non-negative random field)} is a Poisson point process on $D$ with intensity function $\bld$, then $\bX$ is a Cox process driven by $\bld$ \citep[][]{moller2003statistical}. To detail $\bld$ in our context, we specify inhomogeneous NSCP and LGCP in Sections~\ref{sec:NSCP}-\ref{sec:LGCP} and discuss how these models fit earthquake occurrences.
	
	The intensity and pair correlation functions of a Cox process are
	\begin{align*}
		\lambda(u;\bbeta)=\mathbb{E}\bld(u), &&
		g(u,v;\boldsymbol{\psi})=\mathbb{E}[\bld(u)\bld(u)]/\{\mathbb{E}\bld(u)\}^2,
	\end{align*} where the intensity (resp. pair correlation) is now regarded as a parametric function of $\bbeta$ (resp. $\boldsymbol{\psi}$) which will be detailed in the next sections.
	
	\subsection{Neyman-Scott Cox processes} \label{sec:NSCP}
	The Neyman-Scott Cox process model is constructed by generating unobserved mainshocks located according to the stationary Poisson point process. Then each of the mainshocks triggers a random number of aftershocks distributed around the mainshock with a specific spatial probability density function.
	
	In a more technical definition, suppose $\mathbf{C}$ is a stationary Poisson process (mainshock process) with intensity $\kappa>0$. Given $\mathbf{C}$, let $\mathbf{X}_c, c \in \mathbf{C}$, be independent Poisson processes (aftershock processes) with intensity function
	\begin{align*}
		\lambda_c(u;\bbeta)=\exp (\zeta + \bbeta^\top \mathbf{z}(u)) k (u-c; \omega) ,
	\end{align*}
	where $k$ is a probability density function determining the distribution of aftershock points around the mainshocks. Then $\mathbf{X}= \cup_{c \in \mathbf{C}} \mathbf{X}_c$ is an inhomogeneous Ney-man-Scott point process with mainshock process $\mathbf{C}$ and aftershock processes $\mathbf{X}_c, c \in \mathbf{C}$ \citep[e.g.][]{waagepetersen2007estimating,choiruddin2018convex}. The point process $\mathbf{X}$ is a Cox process with intensity function
	\begin{align} \label{eq:intnscp}
		\lambda(u;\bbeta)=\kappa \exp (\zeta + \bbeta^\top \mathbf{z}(u))=\exp (\beta_0 +\bbeta^\top \mathbf{z}(u))
	\end{align}
	driven by random intensity
	\begin{align*}
		\bld(u)=\exp (\zeta+\bbeta^\top \mathbf{z}(u)) {\sum_{c \in \mathbf{C}} k(u-c, \omega})
	\end{align*}
	where $\beta_0=\zeta+\log \kappa$ represents the intercept parameter, $\mathbf{z}(u)=\{z_1(u),\ldots,z_p(u)\}^\top$ is a covariates vector containing $p$-geological variables and $\bbeta=\{\beta_1,\ldots,\beta_p\}^\top$ is the corresponding $p$-dimensional parameter. Note that the additional component $\exp (\zeta + \bbeta^\top \mathbf{z}(u))$ in \eqref{eq:intnscp} accounts for geological variables effect.
	
	Inhomogeneous NSCP could be considered as a natural model for major earthquakes as the superposition of $\mathbf{X}_c$ represents the whole great aftershocks (M $\geq 5$) due to mainshock activity and inhomogeneity due to environmental factors. To determine the distribution of aftershocks around mainshocks, we consider three NSCP models presented in Sections~\ref{sec:thomas}-\ref{sec:vargam}. The first model is the popular Thomas process. The other twos are the Cauchy and variance-Gamma cluster processes as competitors of Thomas process for earthquake modeling (see Sections~\ref{sec:cauchy}-\ref{sec:vargam}).
	
	\subsubsection{Thomas cluster process} \label{sec:thomas}
	The aftershock events are scattered around mainshock locations according to bivariate Gaussian distribution $\mathcal{N}(\mathbf{0},\omega^2 \mathbf{I}_2)$ \citep{moller2003statistical}. The density $k$ is
	\begin{align*}
		k(u;\omega)=(2 \pi \omega^2)^{-1} \exp(-\|u\|^2/(2 \omega^2)).
	\end{align*}
	Smaller values of $\omega$ correspond to tighter clusters and smaller values of $\kappa$ correspond to a fewer number of mainshocks. The parameter vector $\boldsymbol \psi=(\kappa,\omega)^\top$ is referred to as the interaction parameter as it modulates the spatial interaction among earthquake occurrences. The interaction parameter $\boldsymbol \psi$ could be represented by the pair correlation and $K$-functions. The pair correlation function of a Thomas cluster process is
	\begin{align*}
		g(r;\boldsymbol \psi) = 1 + \frac{1}{{4\pi \kappa {\omega ^2}}}\exp \left( { - \frac{{{r^2}}}{{4{\omega ^2}}}} \right), && r=\|u-v\|,
	\end{align*}
	while the $K$-function is
	\begin{align*}
		K(r;\boldsymbol \psi) = \pi {r^2} + \frac{1}{\kappa }\left\{ {1 - \exp \left( { - \frac{{{r^2}}}{{4{\omega ^2}}}} \right)} \right\}.
	\end{align*}
	
	\subsubsection{Cauchy cluster process} \label{sec:cauchy}
	An alternative is to consider a bivariate Cauchy density \citep{jalilian2013decomposition} which has a heavy-tailed distribution so that the aftershock events might be exceedingly distant from the mainshock. The density $k$ is
	\begin{align*}
		k(u;\omega) = \frac{1}{{2\pi {\omega ^2}}}{\left( {1 + \frac{{{{\left\| u \right\|}^2}}}{{{\omega ^2}}}} \right)^{ - \frac{3}{2}}}.
	\end{align*}
	The pair correlation and $K$-functions of the Cauchy cluster process \citep{ghorbani2013cauchy} are respectively
	\begin{align*}
		g(r;\boldsymbol \psi) & = 1 + \frac{1}{{8\pi \kappa {\omega ^2}}}{\left( {1 + \frac{{{{\| r \|}^2}}}{{4{\omega ^2}}}} \right)^{ - \frac{3}{2}}},\\
		K(r;\boldsymbol \psi) & = \pi {r^2} + \frac{1}{\kappa }\left( {1 - \frac{1}{{\sqrt {1 + \tfrac{{{r^2}}}{{4{\omega ^2}}}} }}} \right).
	\end{align*}
	
	\subsubsection{Variance-Gamma cluster process} \label{sec:vargam}
	A variance-Gamma density \citep{jalilian2013decomposition} is of the form
	\begin{align*}
		k(u;\omega) = \frac{1}{{{2^{q + 1}}\pi \omega^2 \Gamma (q + 1)}}\left(\frac{{\left\| u \right\|}}{{\omega }}\right)^q{B_q}\left( {\frac{{\left\| u \right\|}}{\omega}} \right),
	\end{align*}
	where $\Gamma$ is the Gamma function, ${B}$ is a modified Bessel function of the second kind of order $q$ and $q>-1/2$. We fix $q = -1/4$ and treat $q$ as a fixed parameter following the default option of the $\texttt{spastat R}$ package \citep{baddeley2015spatial}. The function $B$ is in particular
	\begin{align*}
		{B}(a) = \frac{{\exp (-a)}}{{\sqrt {2a/\pi } }}\left( {1 + O(1/a)} \right).
	\end{align*}
	The pair correlation and $K$-functions of the variance-Gamma cluster process are respectively
	\begin{align*}
		g(r;\boldsymbol\psi) &= 1 + \frac{1}{{4\pi \kappa \omega^2 q}} {\frac{{{{\left( {{{{\left\| r \right\|}} \mathord{\left/
										{\vphantom {{{{\left\| r \right\|}^2}} \omega }} \right.
										\kern-\nulldelimiterspace} \omega }} \right)}^q}{B_q}\left( {{{\|r\|} \mathord{\left/
								{\vphantom {{\left\| r \right\|} \omega }} \right.
								\kern-\nulldelimiterspace} \omega }} \right)}}{{{2^{q - 1}}\Gamma (q) }}}, \\
		K(r;\boldsymbol \psi) &= \int\limits_0^r {2\pi s g(s;\boldsymbol\psi)ds}.
	\end{align*}

	\subsection{Log-Gaussian Cox Processes} \label{sec:LGCP}
	
	Suppose that $\mathbf{\Lambda}(u)=\exp(\mathbf{G}(u))$, where $\mathbf{G}$ is a Gaussian random field. If conditionally on $\mathbf{\Lambda}$, the point process $\mathbf{X}$ follows the Poisson process, then $\mathbf{X}$ is said to be a log-Gaussian Cox process (LGCP) driven by $\mathbf{\Lambda}$ \citep{moller2003statistical}.
	
	The random intensity $\mathbf{\Lambda}$ could represent a random environment forming aftershock sequences that could be decomposed into different sources of variation with natural interpretation. More precisely, we specify in this paper the $\log \mathbf{\Lambda}$ by
	\begin{align*}
		\log \mathbf{\Lambda}(u)=\zeta + \bbeta^\top \mathbf{z}(u)+\boldsymbol \phi(u),
	\end{align*}
	where the first term is an intercept, the second term represents the geological effects and the last term accounts for the unobserved sources of aftershocks variation. In particular, $\boldsymbol \phi$ is a zero-mean stationary Gaussian random field with covariance function $c(u,v;\boldsymbol \psi)=\sigma^2\exp (-\|u-v\| / \alpha)$. Here, $\boldsymbol \psi=(\sigma^2,\alpha)^\top$ constitutes the interaction parameter vector, where $\sigma^2>0$ is the variance and $\alpha>0$ is the correlation scale parameter.
	
	The intensity function of this LGCP $\mathbf{X}$ is
	\begin{align*}
		\lambda(u;\bbeta)& =\exp(\zeta + \bbeta^\top \mathbf{z}(u) + \sigma^2/2) \\
		& =\exp(\beta_0 + \bbeta^\top \mathbf{z}(u)),
	\end{align*}
	with pair correlation function and $K$-function
	\begin{align*}
		g(r;\boldsymbol \psi) &= \exp ({\sigma ^2}\exp ( - r/\alpha )) \\
		K(r;\boldsymbol \psi) &= \int\limits_0^r {2\pi s\exp ({\sigma ^2}\exp ( - r/\alpha ))ds},
	\end{align*}
	where $\beta_0=\zeta + \sigma^2/2$ plays the role of a new intercept.
	
	\subsection{Parameter estimation and model selection} \label{sec:est}
	We consider two-step procedure \citep{waagepetersen2009two} to estimate $\bbeta$ and $\boldsymbol{\psi}$, where $\hat \bbeta$ is obtained in the first step by maximizing the first-order composite likelihood (Section~\ref{CL}) and $\boldsymbol{\psi}$ is estimated using either maximum second-order composite likelihood or palm likelihood in the second step (Section~\ref{CL2}). The estimation procedure is conducted using the $\texttt{kppm}$ function of the $\texttt{spatstat R}$ package \citep{baddeley2015spatial}. Model selection is performed using the Akaike information criterion (AIC) and envelope test (Section~\ref{sec:sel}) by applying the $\texttt{logLik.kppm}$ and $\texttt{envelope.kppm}$ functions.
	
	\subsubsection{Procedure to estimate intensity parameters} \label{CL} 
	
	The first-order composite likelihood function for $\bbeta$ \citep{waagepetersen2007estimating, moller2007modern} is
	\begin{eqnarray}\label{eq:est2}
		\mathrm{CL}_1(\bbeta)={\sum_{u \in \mathbf{X}  \cap D} \log\lambda(u;\bbeta)} - {\int_D \lambda(u;\bbeta)\mathrm{d}u},
	\end{eqnarray}
	which coincides with the likelihood function of a Poisson point process. To maximize \eqref{eq:est2}, we employ the  Bermann-Turner scheme \citep{berman1992approximating} by discretizing the integral term in \eqref{eq:est2} as
	\begin{align*}
		{\int_{D}  \lambda(u; \bbeta) \mathrm{d}u} \approx {\sum_{i=1}^{M} w(u_i) \lambda (u_i; \bbeta)},
	\end{align*}
	where $u_i, i=1,\ldots,M$ are points in $D$ consisting of the $m$ data points and $M-m$ dummy points and where the $w(u_i)>0$ are quadrature weights such that ${\sum_i w(u_i)}=|D|$. Using this technique, \eqref{eq:est2} is then approximated by
	\begin{align}
		\label{eq:appx:pois}
		\mathrm{CL}_1(\bbeta) \approx \tilde{\mathrm{CL}}_1(\bbeta) = {\sum_{i=1}^{M} w_i \{y_i \log \lambda_i(\bbeta) - \lambda_i(\bbeta)\}},
	\end{align}
	where $w_i=w(u_i), y_i=w_i^{-1} \mathbf{1}(u_i \in \bX \cap D)$ and $\lambda_i(\bbeta)=\lambda (u_i; \bbeta)$.
	Since the equation \eqref{eq:appx:pois} is equivalent to the weighted likelihood function of independent Poisson variables $y_i$ with weights $w_i$, we could apply the standard estimation procedure for Poisson generalized linear models.
	
	\subsubsection{Procedure to estimate cluster parameters} \label{CL2} 
	
	The computationally efficient method to estimate $\boldsymbol{\psi }$ is to maximize the second-order composite likelihood of $\boldsymbol{\psi}$ \citep{guan2006composite,baddeley2015spatial}
	\begin{align}\label{eq:est5}
		\mathrm{CL}_2(\boldsymbol{\psi})  = & \sum_{u,v \in \bX}^{\ne} \mathbf{1}\{ {\| {u - v}\| < R} \}
		\Big\{\log g(u,v;\boldsymbol{\psi}) \nonumber \\ &- \log \int_D\int_D \mathbf{1}\{ {\| {u - v}\| < R} \} g(u,v;\boldsymbol{\psi })\text{d}u \text{d}v\Big\},
	\end{align}
	where $R > 0$ is an upper bound on correlation distance model denoted by $\texttt{rmax}/2$ in the $\texttt{spatstat}$ package. Another alternative is to maximize the palm likelihood defined by
	\begin{align}\label{eq:est8}
		\mathrm{PL}(\boldsymbol{\psi}) = & \sum_{u,v \in \bX}^{\ne} \mathbf{1}\{ {\| {u - v}\| < R} \} \log \lambda_{\mathrm{PL}}(u|v;\boldsymbol{\psi }) \nonumber \\ & - \sum_{v \in \bX} {\int_D \mathbf{1}\{ {\| {u - v}\| < R} \}{\lambda_{\mathrm{PL}}}(u|v;\boldsymbol{\psi })\mathrm{d}u},
	\end{align}
	where
	\begin{align*}
		\lambda_{\mathrm{PL}}(u|v;{\boldsymbol{\psi}}) = \frac{\lambda^{(2)}(u,v;\boldsymbol{\psi })}{\lambda (v)} 
	\end{align*}
	is the palm intensity of the point process model at location $u$ given there is a point at $v$, where $u,v \in D$.
	
	\subsubsection{Model interpretation and selection} \label{sec:sel}
	One particular advantage of parametric modeling of the first and second-order intensity functions is in model interpretability. For example, the effect of geological factors could be easily quantified by assessing the $\bbeta$ estimates. In addition, the spatial correlation parameter $\boldsymbol{\psi }$ could determine the clustering effect, e.g. to identify the number of clusters and the distribution of the aftershocks around mainshocks.
	
	We perform model selection using the envelope test and Akaike information criterion (AIC). Envelopes are the critical bounds of the statistical test of a summary function such as $K$-function which validates suitability point pattern data to point process model \citep{baddeley2015spatial}. To simulate envelopes, we first estimate the inhomogeneous $K$-function for the  earthquake data by
	\begin{align} \label{Kinhom}
		\hat K(r) = \frac{1}{W|D|}\sum_{u,v \in \mathbf{X} \cap D}^{\neq} \frac{\mathbf{1} \{\|u-v\| \leq r\}}{\hat \lambda (u)\hat \lambda(v)}e(u,v;r)
	\end{align}
	where $|D|$ is the {Sulawesi-Maluku area}, $W = \frac{1}{|D|}\sum_{u \in \mathbf{X} \cap D}\hat \lambda(u)^{-1}$, and $e(u,v;r)$ is an edge correction. Next we generate $n$ point patterns as realizations of the null model and compute their $K$-function estimates, {denoted by} $\hat{K}^{1}(r), \ldots,\hat{K}^{n}(r)$. The upper and lower envelopes are then obtained by resp. determining the maximum and minimum of the $n$ $K$-function estimates. We reject the null hypothesis with significance level $\gamma=2/(n+1)$ if the estimated inhomogeneous $K$-function \eqref{Kinhom} lies outside the interval bound. The envelope test is useful to determine the most suitable model among list of candidate models.
	
	As an alternative, we compare models by evaluating their {Akaike information criterion (AIC) values and select the model with minimum AIC. The AIC \citep{choiruddin2020information} is defined by}
	\begin{align*}
		\mathrm{AIC}=-2\mathrm{L_{max}}+2p,
	\end{align*}
	where $\mathrm{L_{max}}$ is the maximum of either \eqref{eq:est2} for the Poisson process, \eqref{eq:est5} or \eqref{eq:est8} for the Cox processes, and where $p$ is the number of overall estimated parameters.

	\section{Results} \label{sec:result}
	\subsection{Exploratory data analysis}
	
	It is in general difficult to distinguish between clustering and spatial inhomogeneity due to confounding issues. To investigate both effects in our study, we first study the spatial trend by quadrat counting test and then employ $K$-function plot to detect the interpoint interaction. Such technique is popular in the point pattern analysis \citep{baddeley2015spatial}. Various functions in the $\texttt{spatstat}$ package are applied.
	
	\begin{figure*}[h]
		\centering
		\subfloat[]{\includegraphics[width=0.48\textwidth]{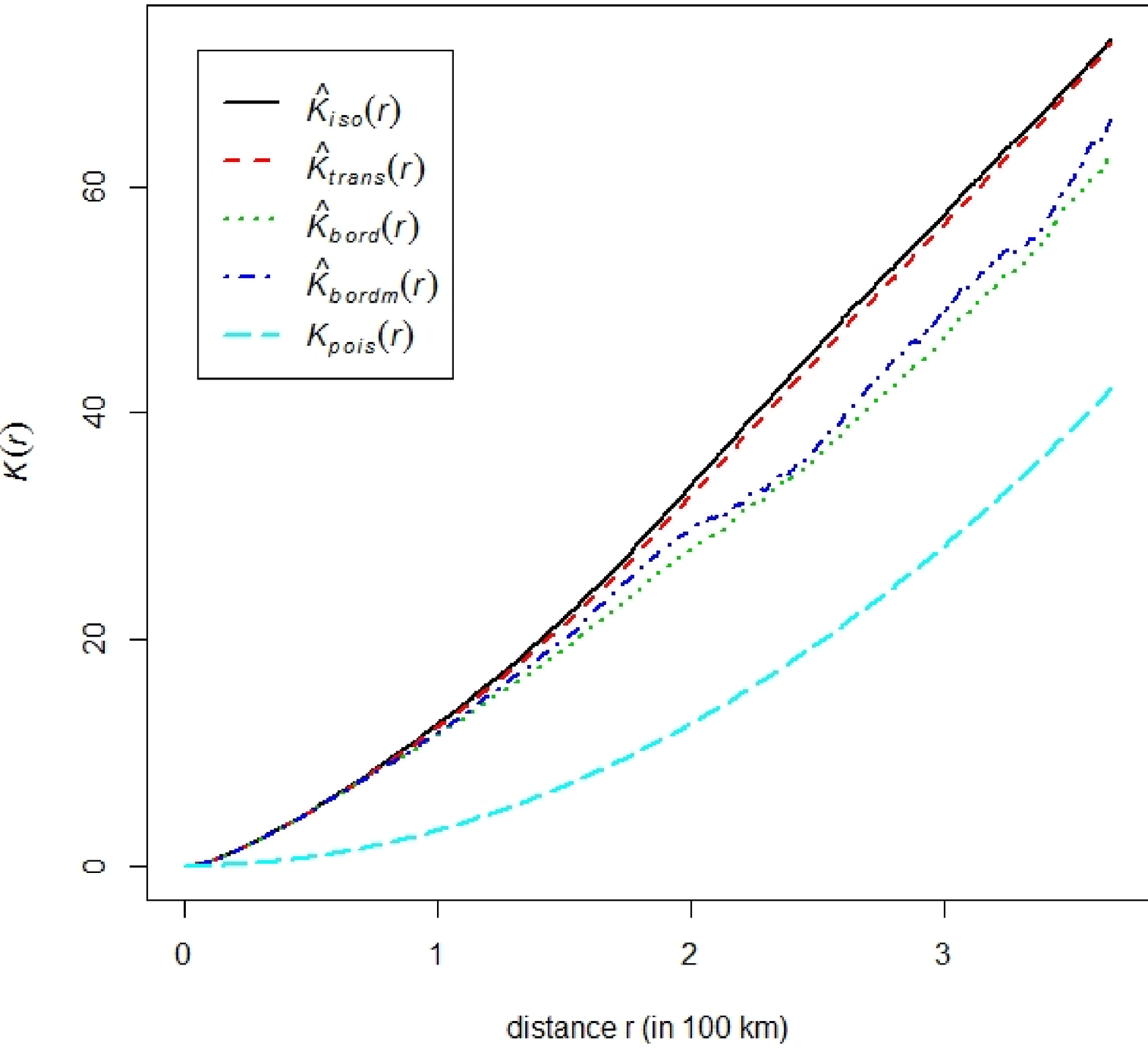}}
		\subfloat[]{\includegraphics[width=0.48\textwidth]{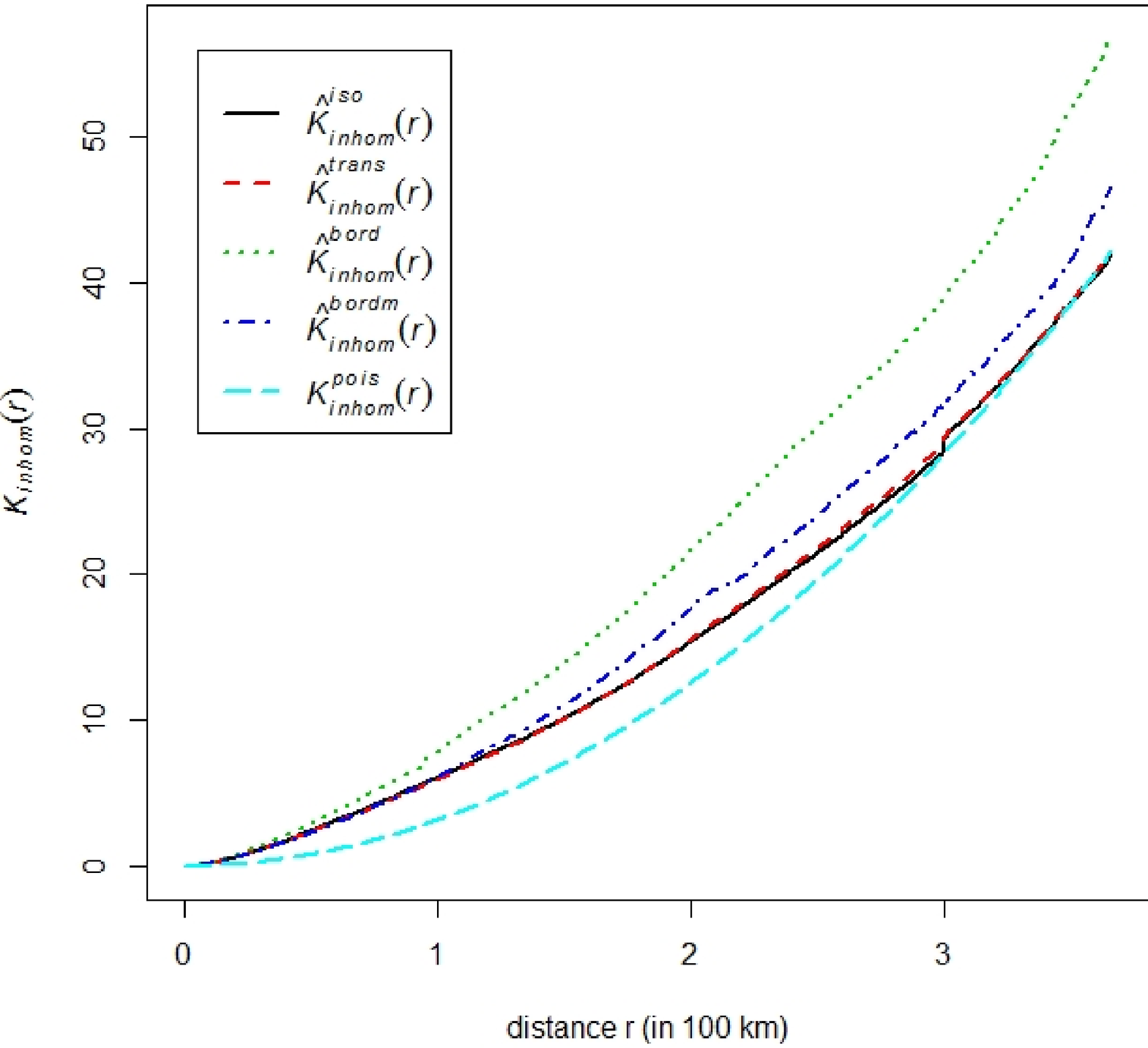}}
		\caption{(a) The $K$-function plot  and (b) inhomogeneous $K$-function plot  for the earthquake dataset in Sulawesi-Maluku with magnitudes of larger than 5 during 2009-2018.}
		\label{kinhom}
	\end{figure*}

	We use the $\texttt{quadrat.test}$ function to apply quadrat counting test with a null hypothesis of stationary Poisson process against an inhomogeneous Poisson process. The main idea is to divide the window $D$ into $l$ sub-windows ${D_1},\ldots,{D_l}$ and test under the Poisson process assumption whether the number of points in each sub-window is significantly different. By dividing the observation domain into $10 \times 10$ grids, the value of the chi-square statistic is 2853.3 with the degree of freedom 99 ($p$-value is equal to $2.2 \times {10^{ - 16}}$). Therefore, under the assumption that the process is Poisson, there is an evidence of spatial trend possibly due to geological factors. Next we consider inhomogeneous $K$-function (Figure~\ref{kinhom}b) to investigate clustering using the $\texttt{Kinhom}$ function. The baby blue dashed line shows the $K$-function of an inhomogeneous Poisson process. In general, the edge corrected inhomogeneous $K$-function for the earthquake data  indicates the presence of clustering especially when the radius is less than 250 km. The {clustering assuming inhomogeneity} (Figure~\ref{kinhom}b) is not as strong as the one of the stationary assumption (Figure~\ref{kinhom}a) since the source of earthquake variations due to the spatial trend has been filtered out. This remarks that assuming stationary point process could be dangerous for the earthquake data as all sources of variability are only covered by clustering (see also Sections~\ref{sec:compare}-\ref{sec:inter}). We investigate in more detail the spatial trend due to geological variables and clustering effect using various inhomogeneous Cox point process models in the next Sections.
	
	\subsection{Inference and model selection}
	
	To model the earthquake epicenters in Sulawesi and Maluku involving geological variables by using the inhomogeneous Thomas, Cauchy, variance-Gamma, and log-Gaussian Cox processes, we consider two-step estimation technique (Section~\ref{sec:est}) by employing the $\texttt{kppm}$ function of the $\texttt{spatstat}$. To compare with the independence case, we also apply the Poisson point process model with intensity $\lambda(u;\bbeta) =\exp(\beta_0 + \sum_{i=1}^3 \beta_i {z_i}(u))$, where $z_i,i=1,2,3$ are detailed by Table~\ref{table:covs}. We maximize the Poisson likelihood \eqref{eq:est2} to obtain $\bbeta$ estimates using the $\texttt{ppm}$ function. We also apply the stationary Poisson and cluster point process models.  We compare all the models in Section~\ref{sec:compare} and discuss the best ones in Section~\ref{sec:inter}.
	
	\subsubsection{Model comparison} \label{sec:compare}
	
	\renewcommand{\arraystretch}{1.25}
	\begin{table*}[htbp!]
    \setlength{\tabcolsep}{2pt}
    \centering\small
    \caption{The clustering estimates $\hat{\boldsymbol{\psi }}$, where $\hat{\boldsymbol{\psi }}=(\hat \kappa, \hat \omega)$ for the NSCP models and $\hat{\boldsymbol{\psi }}=(\hat \sigma^2, \hat \alpha)$ for the LGCP, the maximum composite and palm likelihoods $\mathrm{L_{max}}$ and the AIC values for each of fitted model to the Sulawesi-Maluku earthquake data}
		\resizebox{\textwidth}{!}
        {
			\begin{tabularx}{\textwidth}{l*{9}{r}}
				\toprule
				&&\multicolumn{4}{c}{Second-order composite likelihood}
				&\multicolumn{4}{c}{Palm likelihood}\\\cmidrule(r){3-6}\cmidrule(l){7-10}
				& Poisson& Thomas & Cauchy&Var-Gamma & LGCP & Thomas &Cauchy& Var-Gamma & LGCP \\ \hline
				$\hat \kappa$&-&0.53&0.28&0.40&-&0.13&0.09&0.11&-\\
				$\hat\omega$&-&0.24&0.19&0.28&-&0.49&0.34&0.55&- \\
				$ \hat \sigma^2$&-&-&-&-&1.97&-&-&-&1.98\\
				$\hat \alpha$&-&-&-&-&0.44&-&-&-&0.79\\ \midrule
				
				$\mathrm{L_{max}} \; (\times 10^3)$&0.62&647.54&647.80&647.80&647.67&145.23&146.55&146.64&146.96\\
				AIC $(\times 10^4)$&-0.12&-129.51&-129.56&-129.56&-129.53&-29.04&-29.31&-29.33&-29.39\\
				\bottomrule
				\label{table:est.result}
		\end{tabularx}}
    \end{table*}
	
	\begin{figure*}[h]
		\centering
		\subfloat[]{\includegraphics[width=0.325\textwidth]{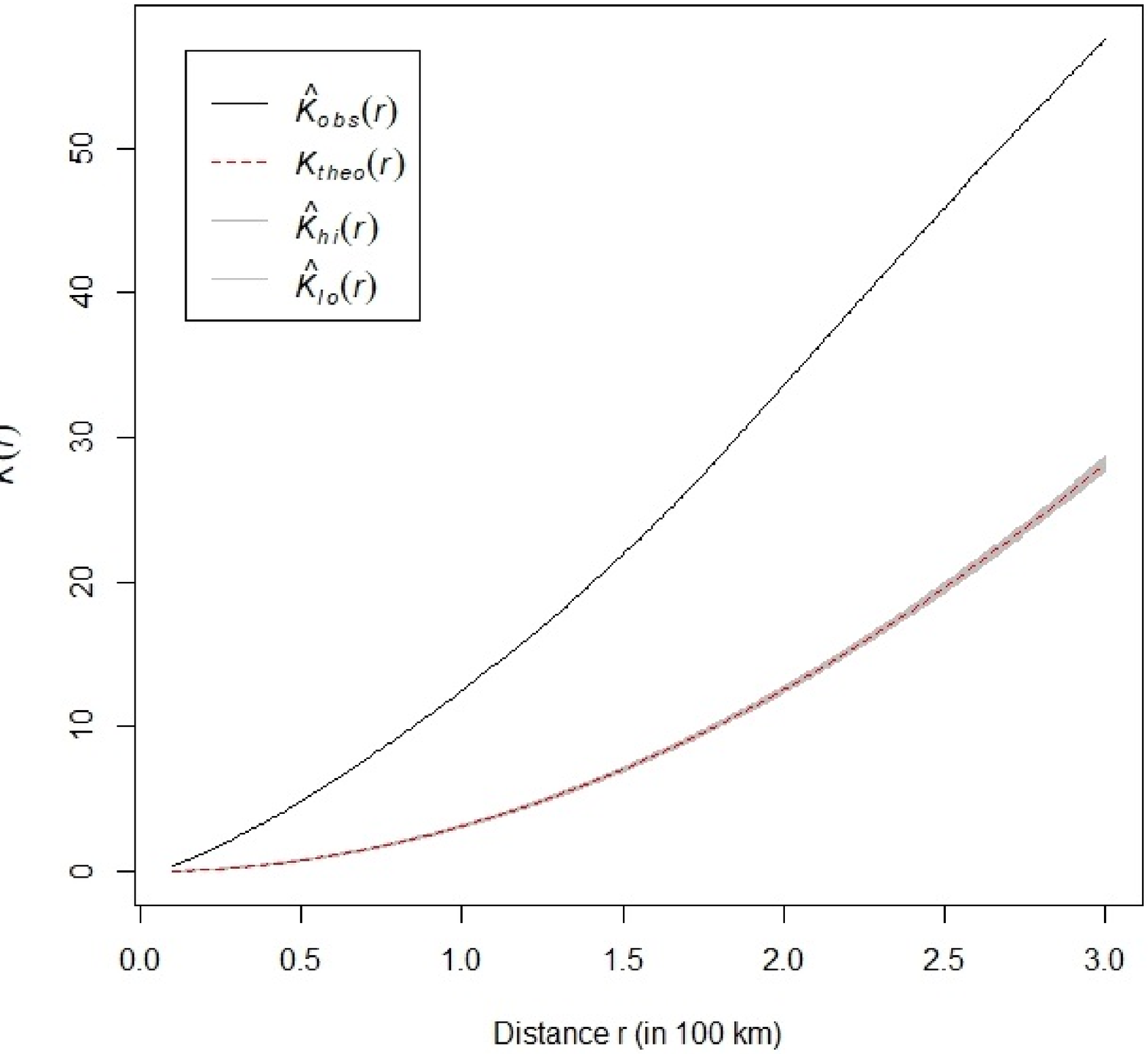}}
		\subfloat[]{\includegraphics[width=0.325\textwidth]{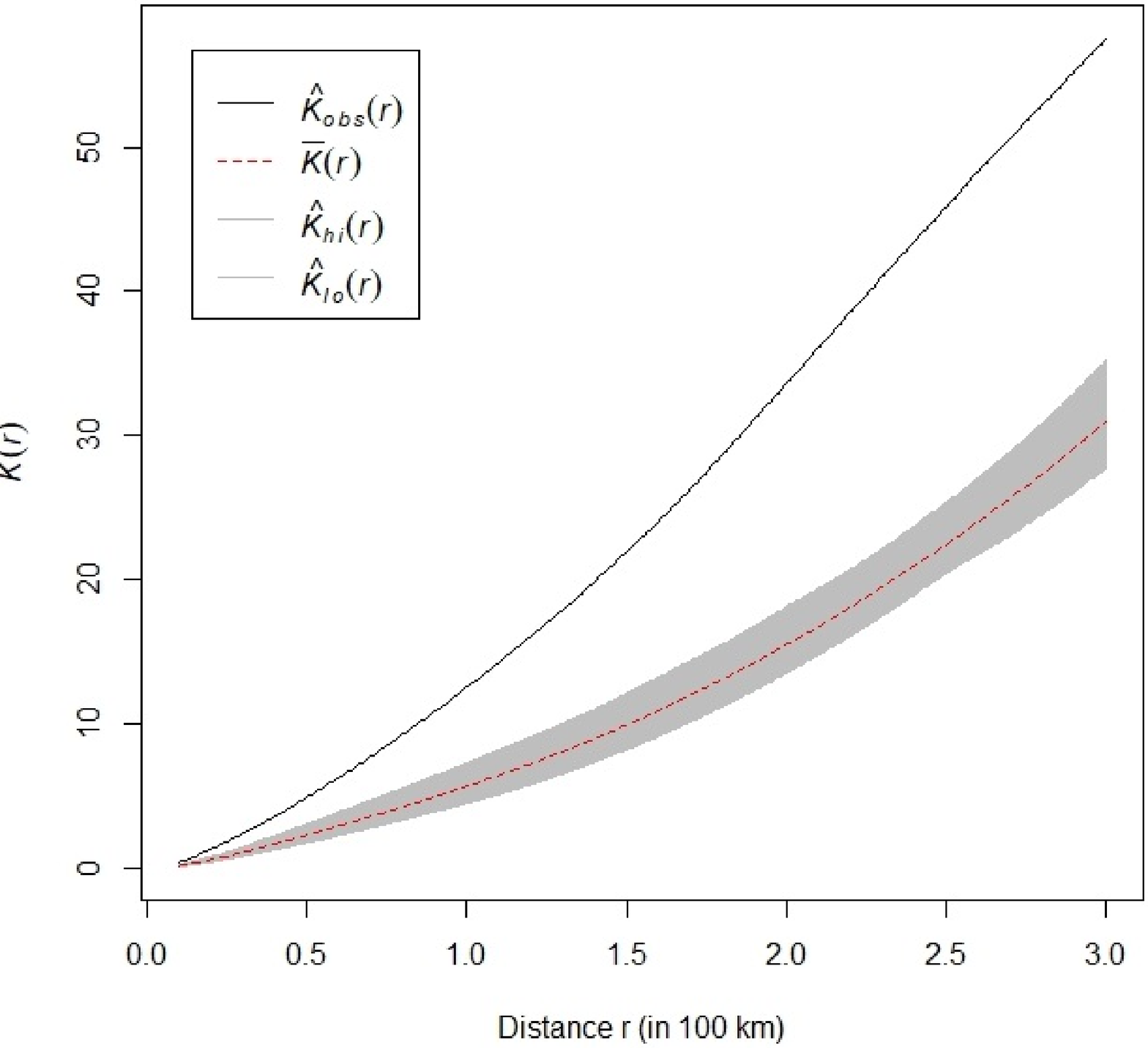}} \subfloat[]{\includegraphics[width=0.325\textwidth]{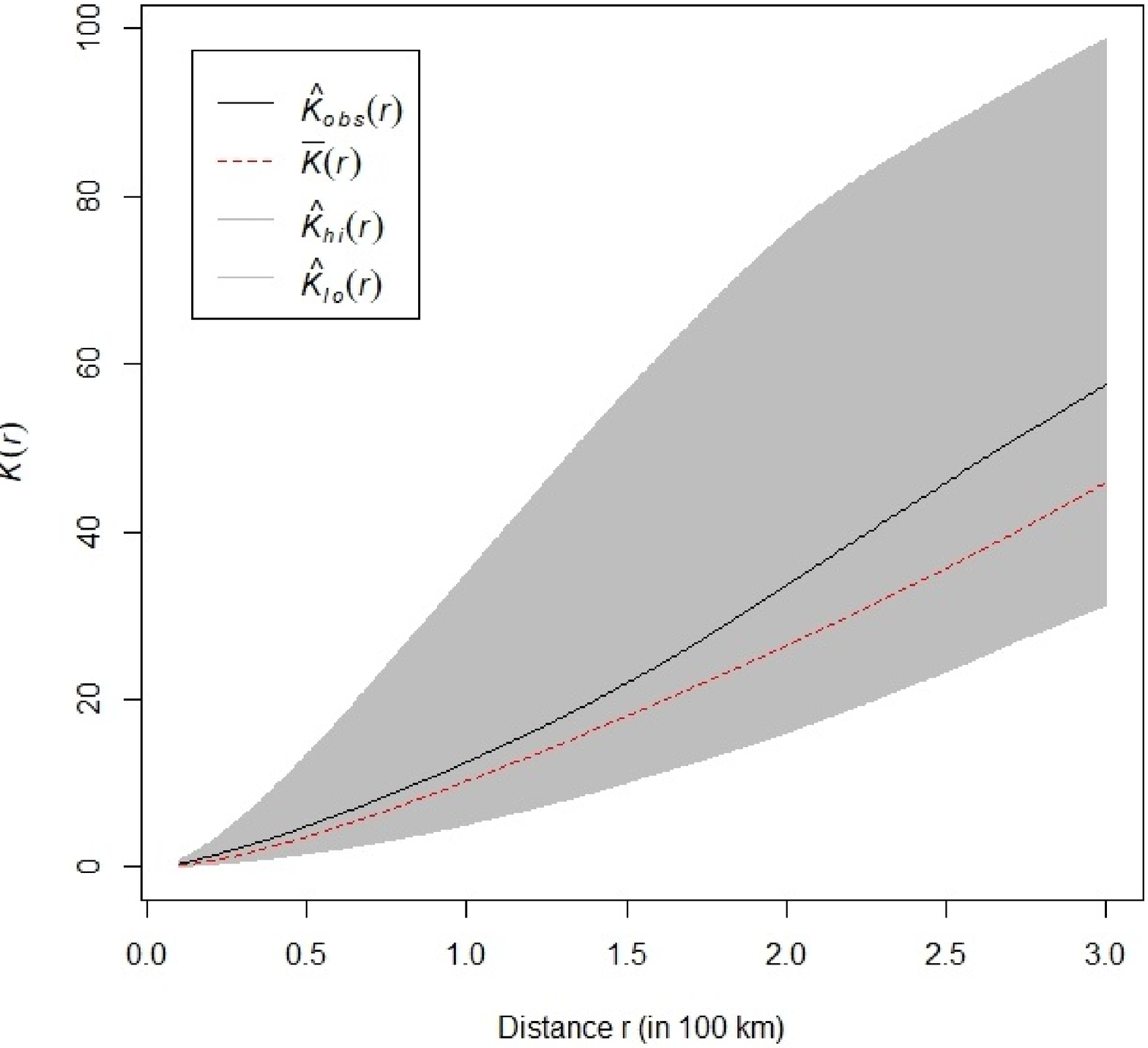}}
		\caption{Envelopes $K$-function for the earthquake data assuming stationary models: (a) Poisson, (b) variance-Gamma with \texttt{clik2}, and (c) LGCP with \texttt{palm}. \texttt{clik2} and \texttt{palm} are the estimation considering respectively the second-order composite and palm likelihoods.}
		\label{env:kppmStas}
	\end{figure*}

	\begin{figure*}
		\centering
		\subfloat[]{\includegraphics[width=0.32\textwidth]{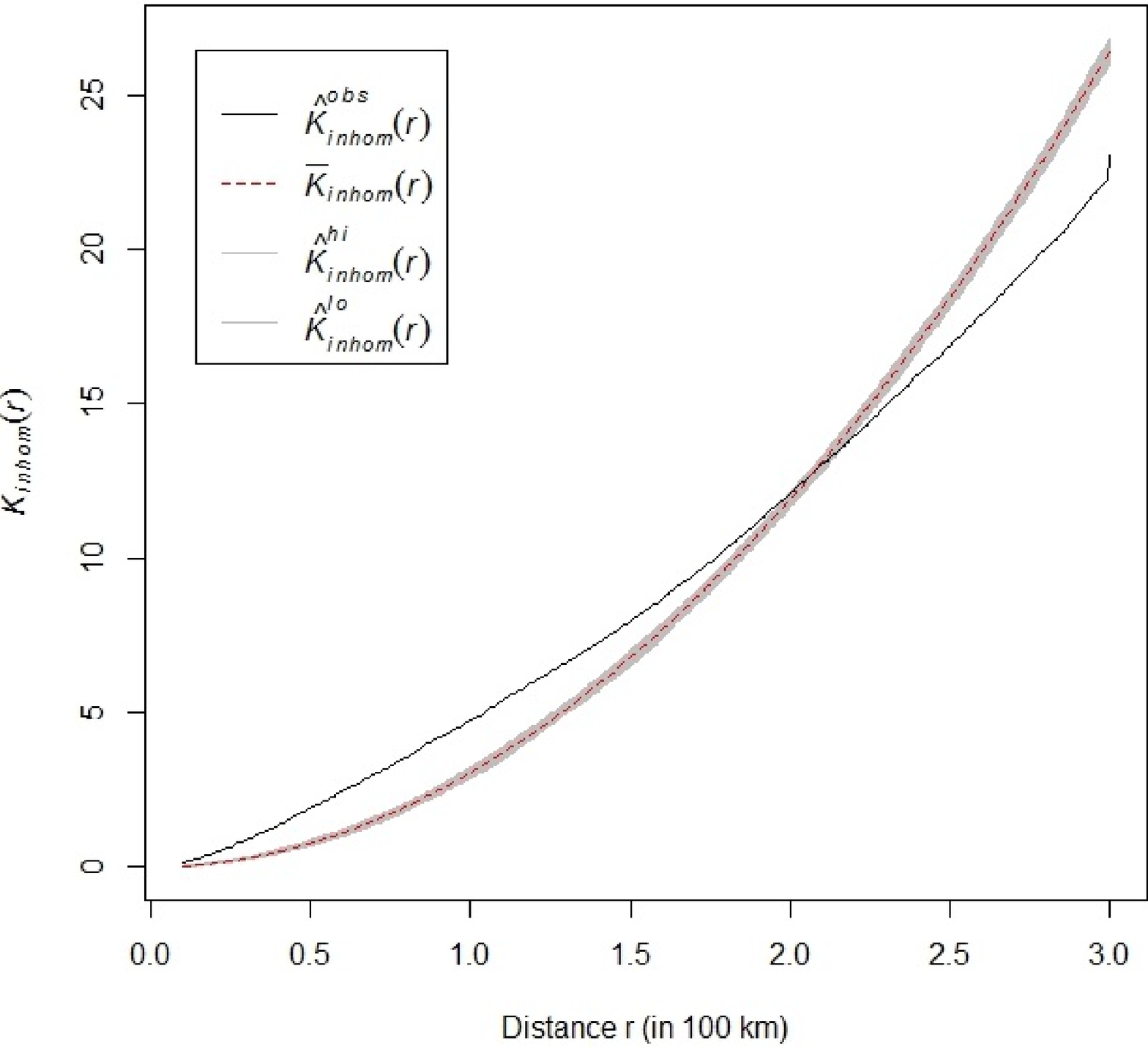}}
		\subfloat[]{\includegraphics[width=0.32\textwidth]{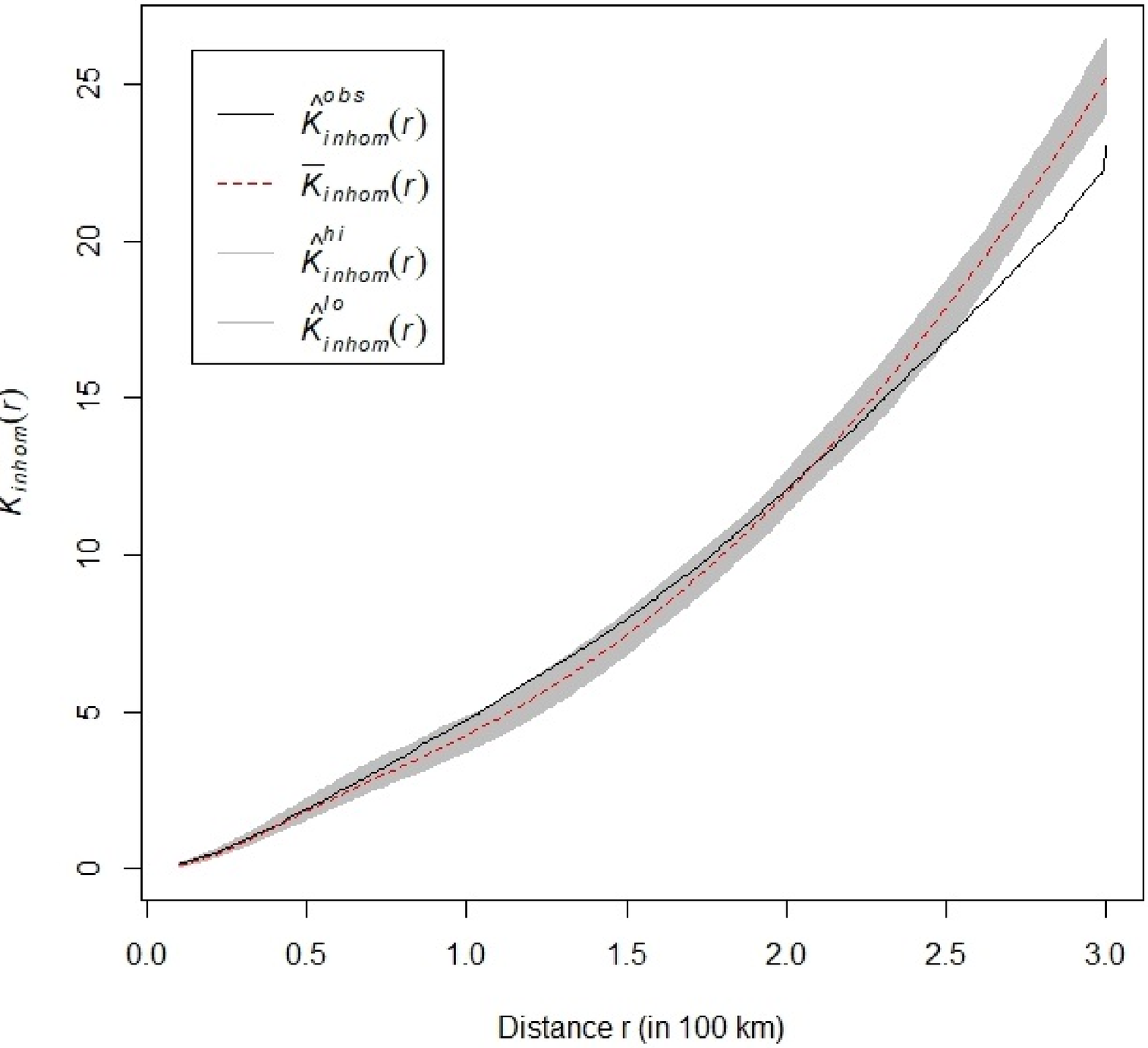}}
		\subfloat[]{\includegraphics[width=0.32\textwidth]{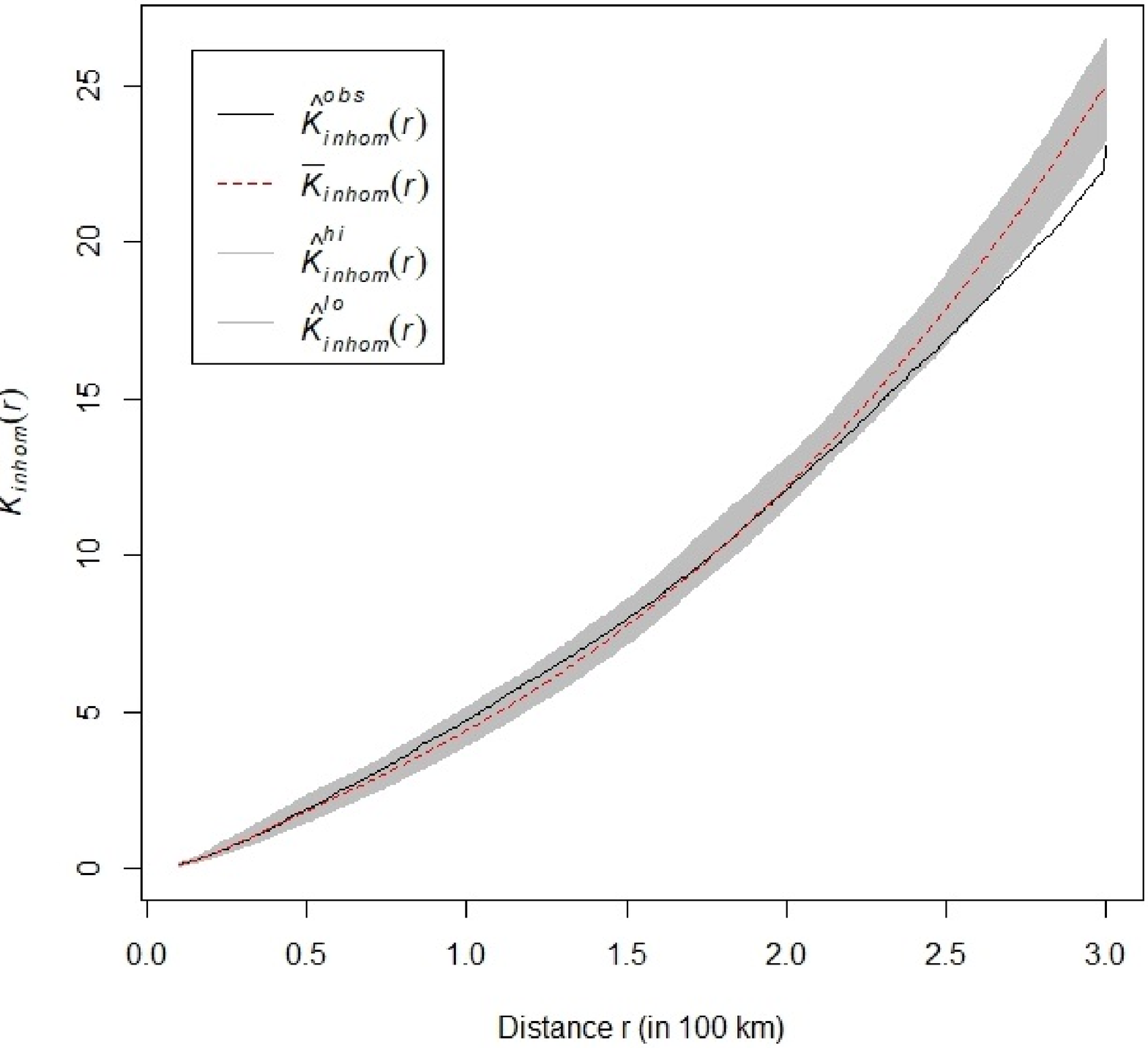}}\\
		\subfloat[]{\includegraphics[width=0.32\textwidth]{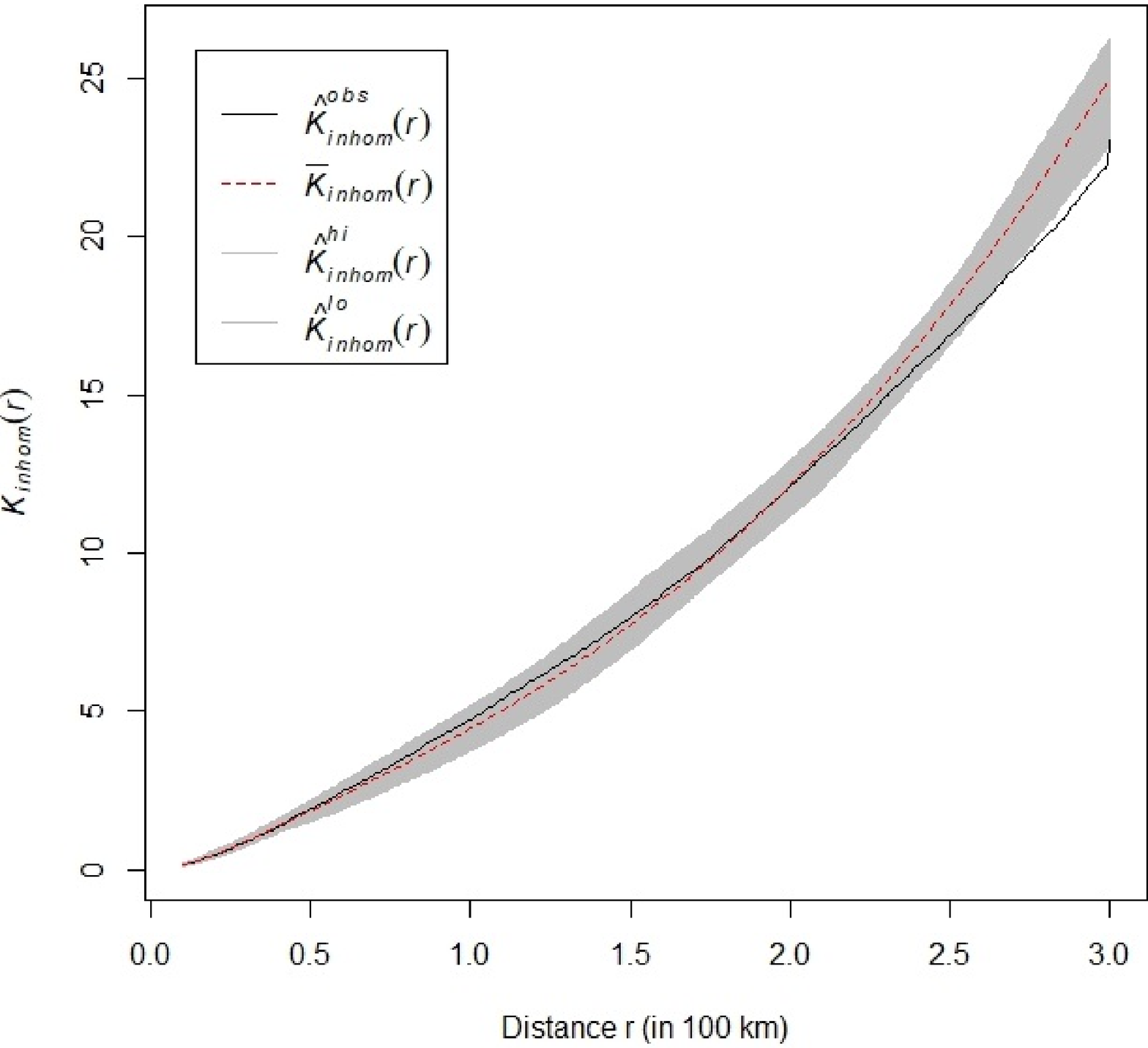}}
		\subfloat[]{\includegraphics[width=0.32\textwidth]{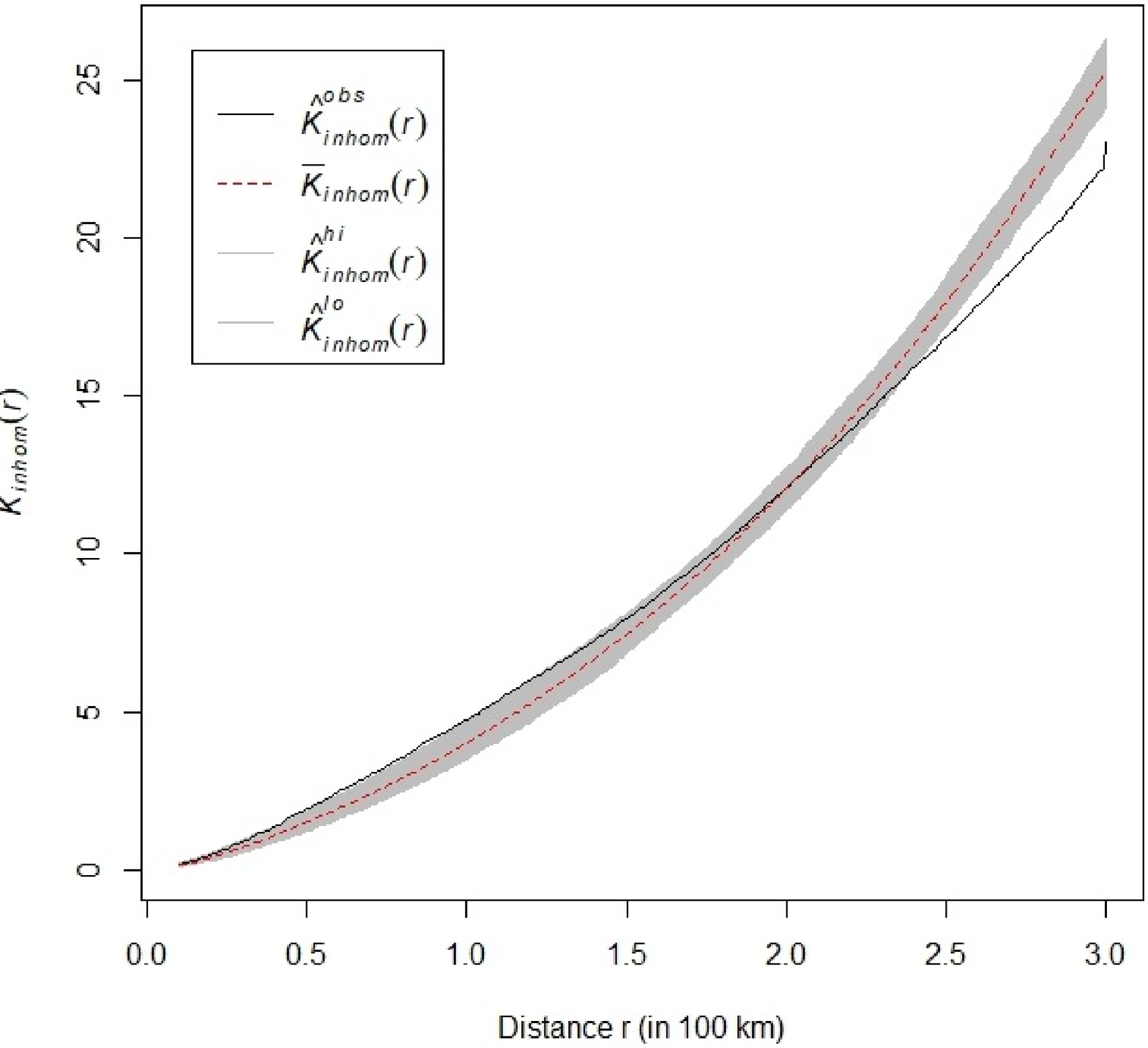}}	\subfloat[]{\includegraphics[width=0.32\textwidth]{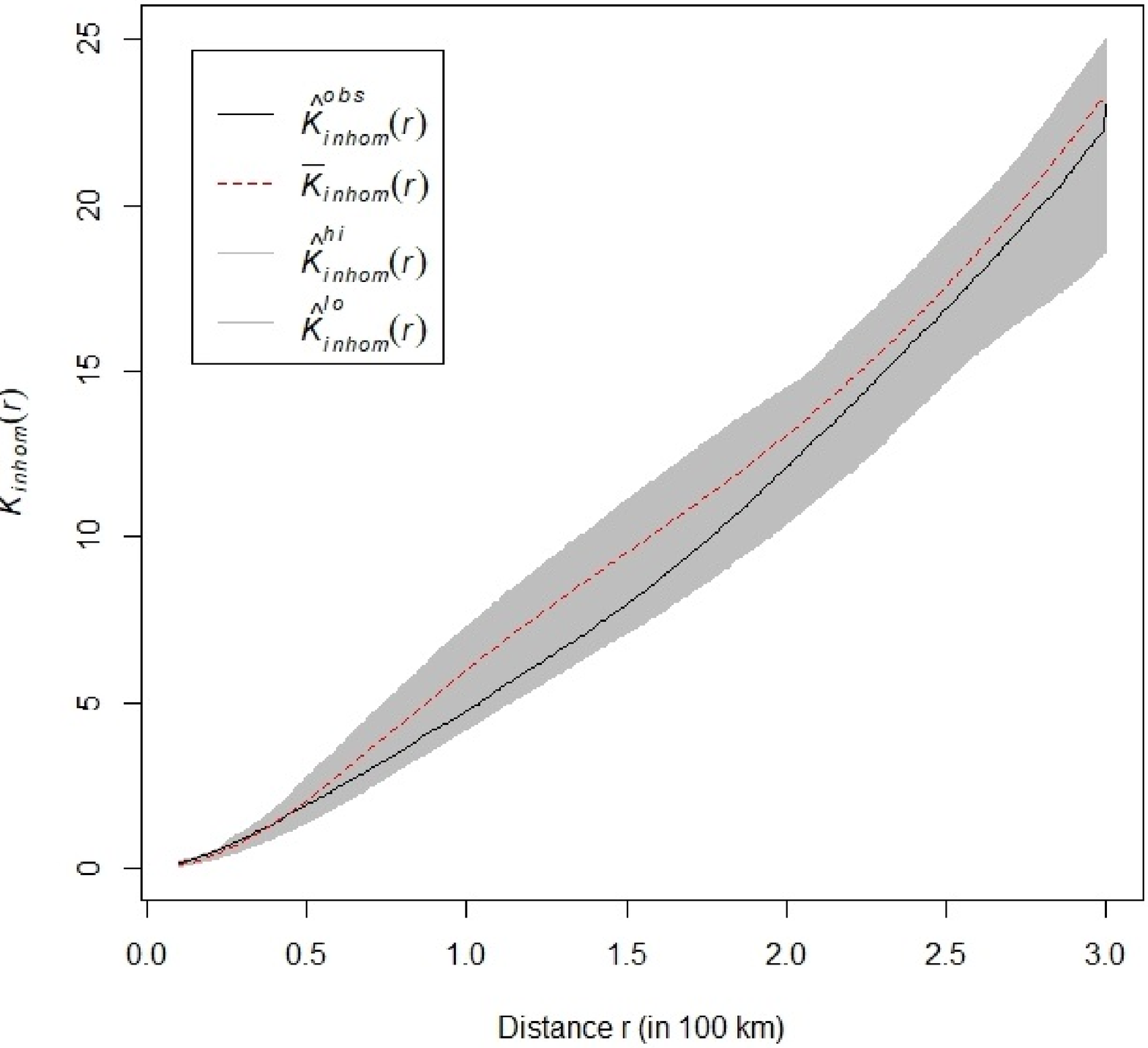}}\\
		\subfloat[]{\includegraphics[width=0.32\textwidth]{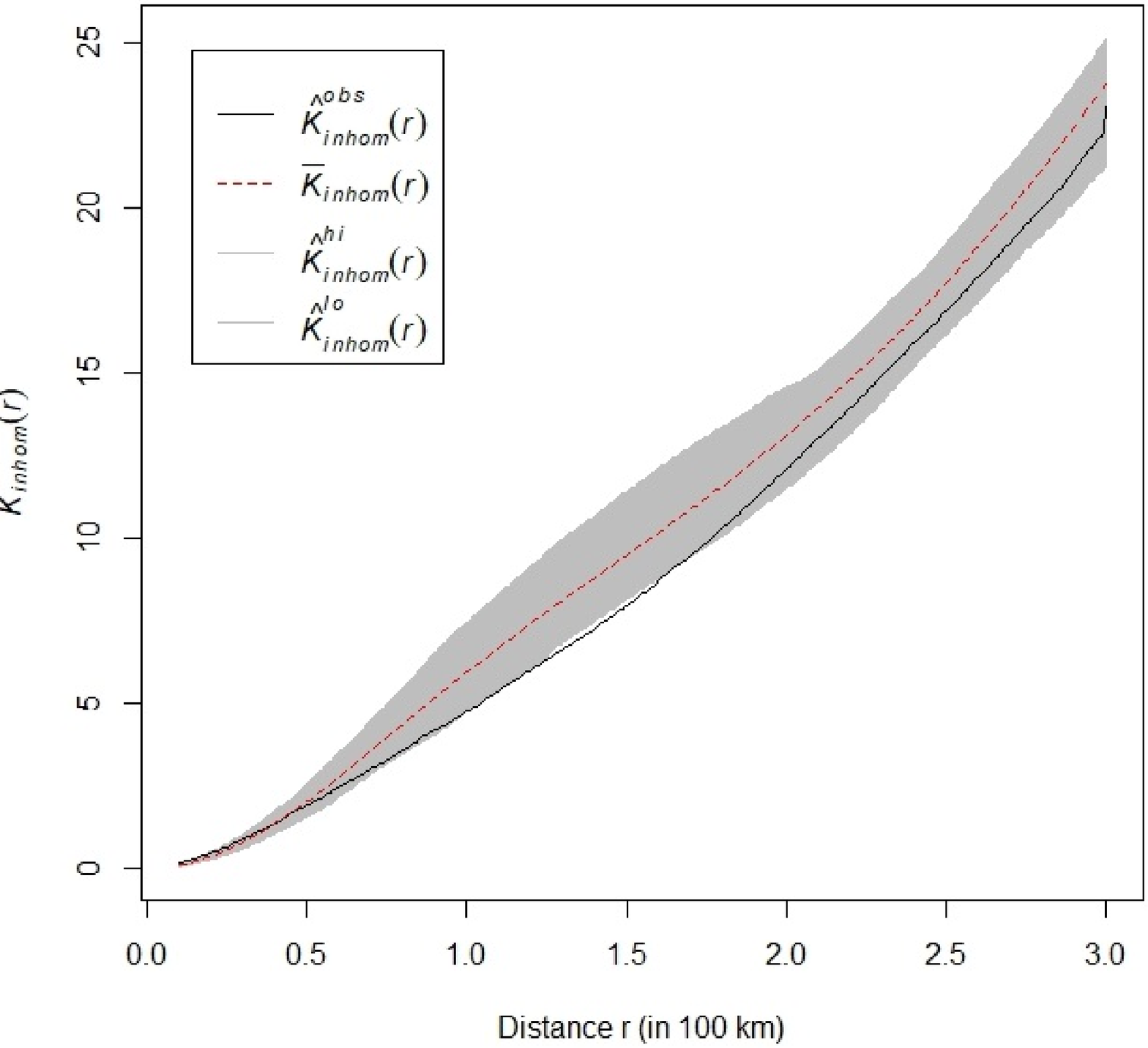}} \subfloat[]{\includegraphics[width=0.32\textwidth]{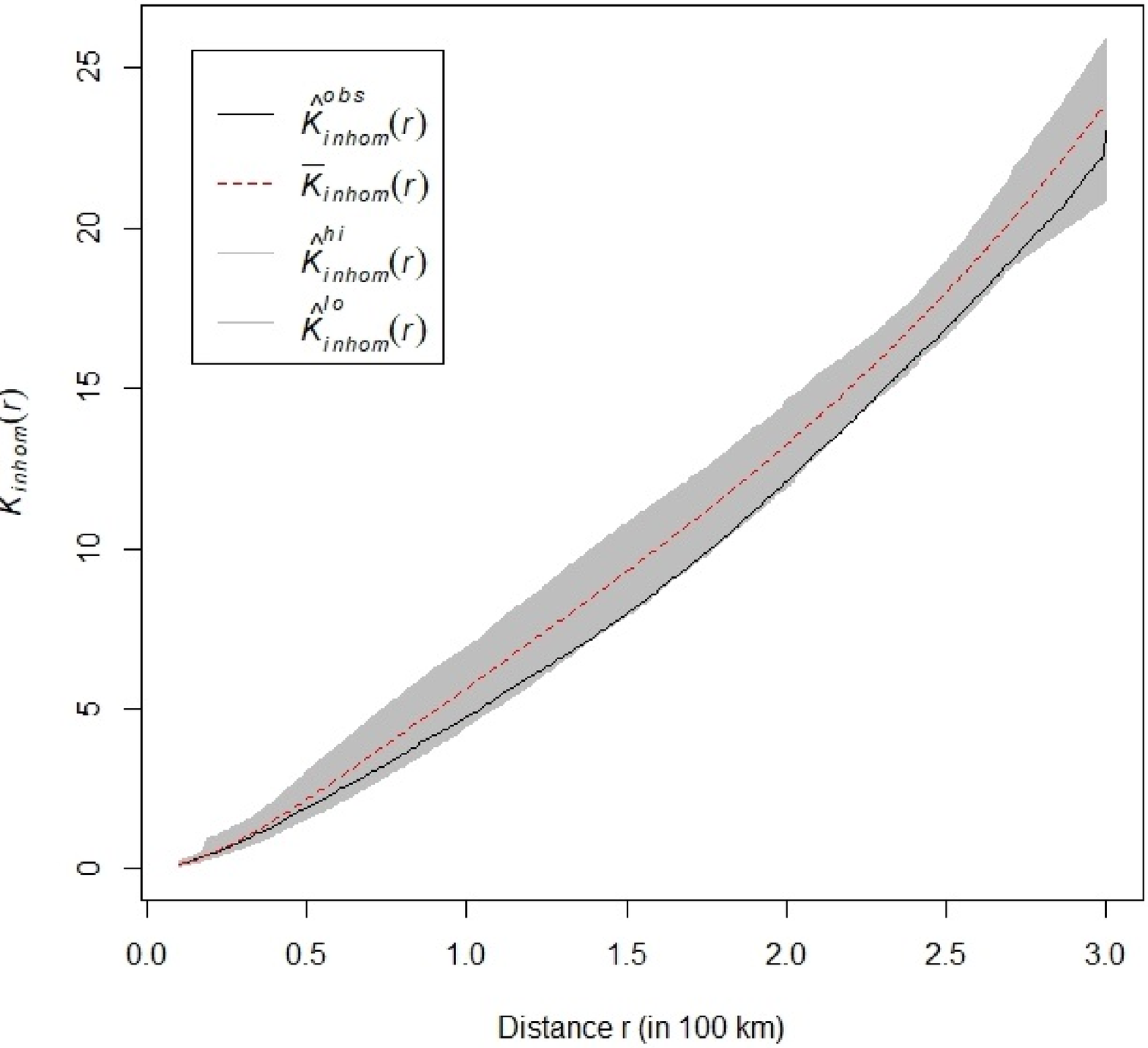}}
		\subfloat[]{\includegraphics[width=0.32\textwidth]{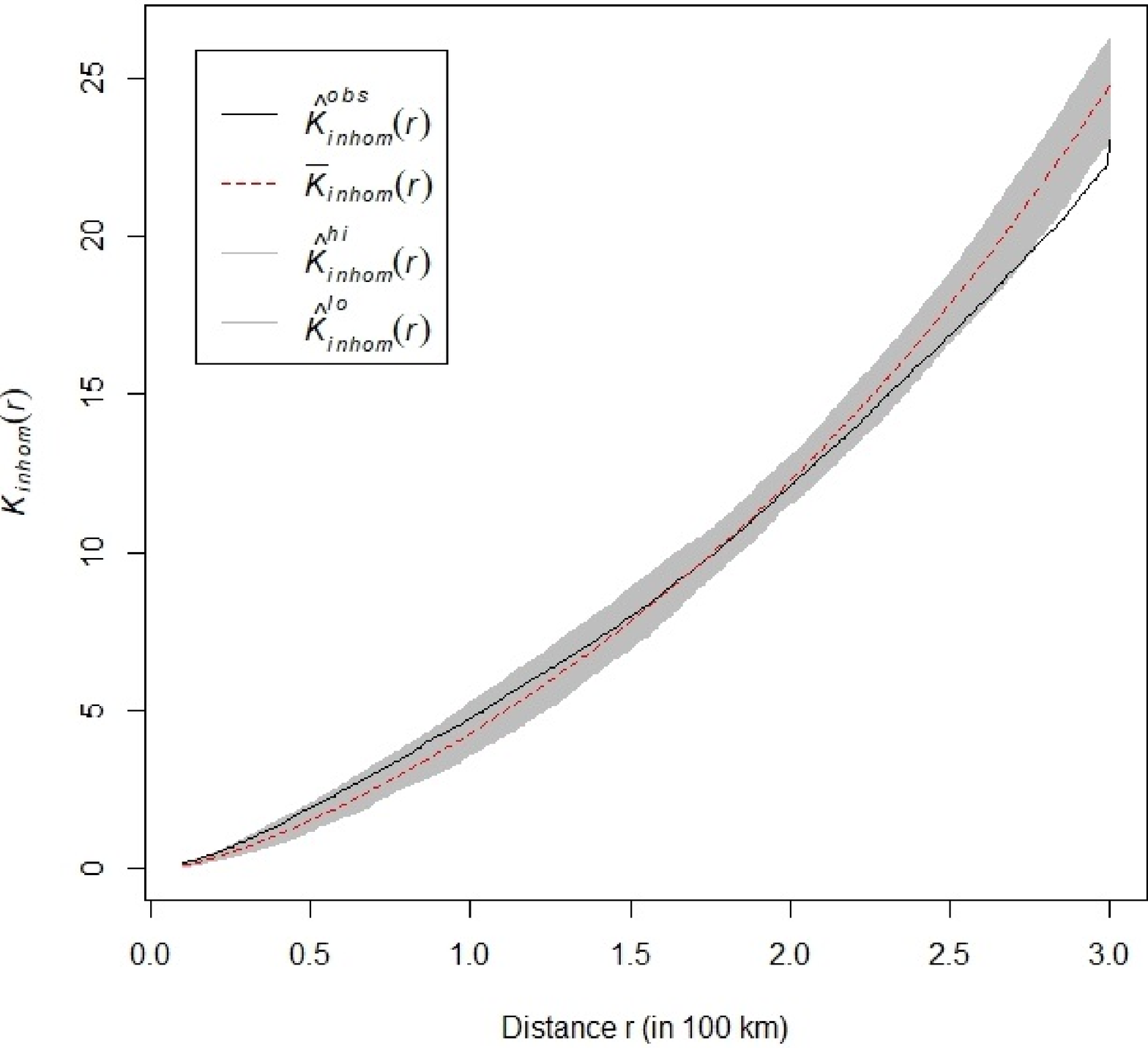}}
		\caption{Envelopes inhomogeneous $K$-function for the earthquake data using: (a) Poisson, (b) Thomas, (c) Cauchy, (d) variance-Gamma, (e) LGCP, (f) Thomas, (g) Cauchy, (h) variance-Gamma, and (i) LGCP. The parameters on (b)-(e) models are estimated by second-order composite likelihood while (f)-(i) are estimated by palm likelihood.}
		\label{env:kppm}
	\end{figure*}

The resulting estimators are presented in Table~\ref{table:est.result}. We only report the estimates which correspond to the correlation parameters, i.e. $\boldsymbol{\psi }$. The $\bbeta$ estimates are identical for all models, so it is not useful for model comparison. The detailed interpretation of the $\bbeta$ estimates is discussed in Section~\ref{sec:inter}.
	
	Concerning the NSCP models, the Cauchy model captures the highest clustering effect (smallest $\hat \kappa$ and $\hat \omega$). Thomas and variance-Gamma models are quite similar, but we notice that the variance-Gamma model allows a larger variance with a smaller intensity of the mainshock process. Regarding different estimation methods, estimation by second-order composite likelihood technique mainly detects the clustering effect by a smaller variance of the aftershock distribution while estimation by palm likelihood detects by a smaller intensity of the mainshock process (smaller $\hat \kappa$). For the LGCP, the estimation using second-order composite likelihood is more clustered as $\hat \alpha$ is approximately twice smaller with a similar variance.
	
	The Poisson model is obviously failed by producing the largest AIC. By two additional cluster parameters, the cluster models outperform the Poisson model. The largest maximum likelihood (and minimum AIC) for the cluster models is obtained when second-order composite likelihood estimation is employed. The similar messages apply when the envelope $K$-function test (Figures~\ref{env:kppmStas}-\ref{env:kppm}) is performed. We in particular notice that estimation using palm likelihood (Figures~\ref{env:kppm}f-i) results in a wider interval.  We also display in Figure~\ref{env:kppmStas} the envelope test under an assumption that the process is stationary. Apart from the stationary Poisson process model, we select stationary variance-Gamma cluster and LGCP models which obtain the smallest AIC using respectively the second-order composite and palm likelihood estimation. We observe that stationary models much underfit the observed $K$-function. The popular Thomas model does not perform as well as the Cauchy and variance-Gamma models may be due to the largest estimate of $\kappa$ (overestimate the mainshock intensity). In addition, one of the main features of Cauchy and variance-Gamma models over Thomas model is that the two earlier models allow the aftershock scattered very distant around the mainshock. We conclude based on Table~\ref{table:est.result} and Figures~\ref{env:kppmStas}-\ref{env:kppm} that the inhomogeneous Cauchy and variance-Gamma models with second-order composite likelihood estimation perform best, therefore, we focus to interpret the two models in the next section. We notice that the variance-Gamma produces the smallest AIC, but the one by Cauchy is quite similar. The performance using envelope $K$-function also shows they are pretty similar.

	The resulting estimators are presented in Table~\ref{table:est.result}. We only report the estimates which correspond to the correlation parameters, i.e. $\boldsymbol{\psi }$. The $\bbeta$ estimates are identical for all models, so it is not useful for model comparison. The detailed interpretation of the $\bbeta$ estimates is discussed in Section~\ref{sec:inter}.
	
	Concerning the NSCP models, the Cauchy model captures the highest clustering effect (smallest $\hat \kappa$ and $\hat \omega$). Thomas and variance-Gamma models are quite similar, but we notice that the variance-Gamma model allows a larger variance with a smaller intensity of the mainshock process. Regarding different estimation methods, estimation by second-order composite likelihood technique mainly detects the clustering effect by a smaller variance of the aftershock distribution while estimation by palm likelihood detects by a smaller intensity of the mainshock process (smaller $\hat \kappa$). For the LGCP, the estimation using second-order composite likelihood is more clustered as $\hat \alpha$ is approximately twice smaller with a similar variance.
	
	The Poisson model is obviously failed by producing the largest AIC. By two additional cluster parameters, the cluster models outperform the Poisson model. The largest maximum likelihood (and minimum AIC) for the cluster models is obtained when second-order composite likelihood estimation is employed. The similar messages apply when the envelope $K$-function test (Figures~\ref{env:kppmStas}-\ref{env:kppm}) is performed. We in particular notice that estimation using palm likelihood (Figures~\ref{env:kppm}f-i) results in a wider interval.  We also display in Figure~\ref{env:kppmStas} the envelope test under an assumption that the process is stationary. Apart from the stationary Poisson process model, we select stationary variance-Gamma cluster and LGCP models which obtain the smallest AIC using respectively the second-order composite and palm likelihood estimation. We observe that stationary models much underfit the observed $K$-function. The popular Thomas model does not perform as well as the Cauchy and variance-Gamma models may be due to the largest estimate of $\kappa$ (overestimate the mainshock intensity). In addition, one of the main features of Cauchy and variance-Gamma models over Thomas model is that the two earlier models allow the aftershock scattered very distant around the mainshock. We conclude based on Table~\ref{table:est.result} and Figures~\ref{env:kppmStas}-\ref{env:kppm} that the inhomogeneous Cauchy and variance-Gamma models with second-order composite likelihood estimation perform best, therefore, we focus to interpret the two models in the next section. We notice that the variance-Gamma produces the smallest AIC, but the one by Cauchy is quite similar. The performance using envelope $K$-function also shows they are pretty similar.
	
	\subsubsection{Model interpretation} \label{sec:inter}
	
	\renewcommand{\arraystretch}{1.25}
	\begin{table*}[htbp!]
		\caption{The important spatial covariates and their corresponding regression $\bbeta$ estimates with a significance level 0.05 for the earthquake data. The $\hat \theta$ is either $\hat \beta_1$ or $\hat \beta_2$ }
		{\begin{tabularx}{\textwidth}{l*{4}{>{\centering\arraybackslash}X}}
				\toprule
				&$\hat \theta$&$\exp(\hat \theta)$&$1/\exp{\hat \theta}$\\ \midrule
				$z_1$&-0.363&0.696&1.437\\
				$z_2$&-0.276&0.759&1.318\\
				\bottomrule
				\label{table:varGam2cov}
		\end{tabularx}}
	\end{table*}
	
	\begin{figure*}[h]
		\centering
		\subfloat[]{\includegraphics[width=0.49\textwidth]{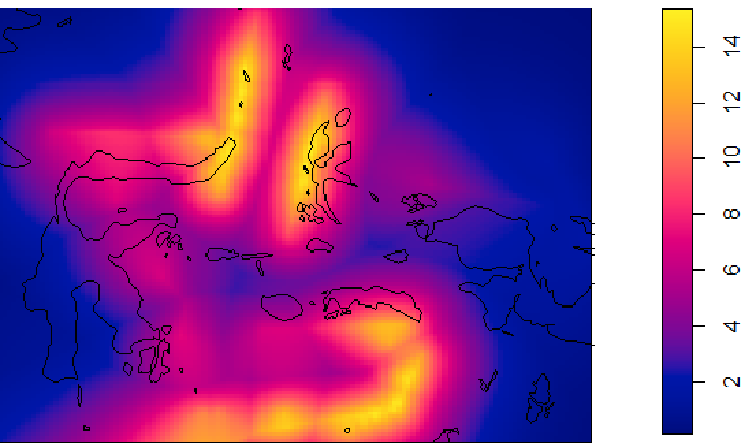}}
		\subfloat[]{\includegraphics[width=0.49\textwidth]{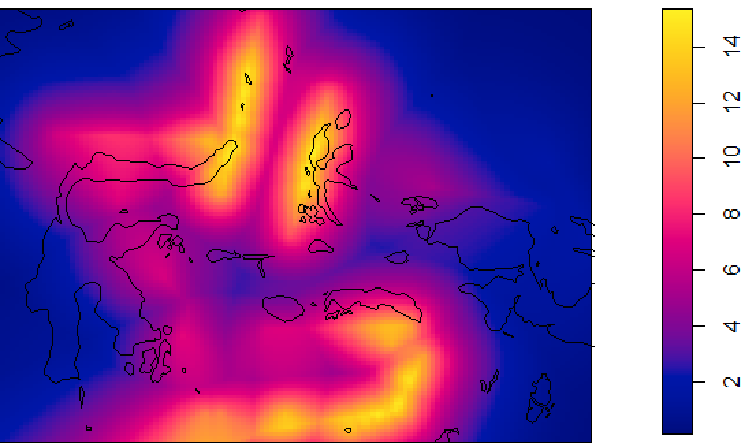}} \\
		\subfloat[]{\includegraphics[width=0.65\textwidth]{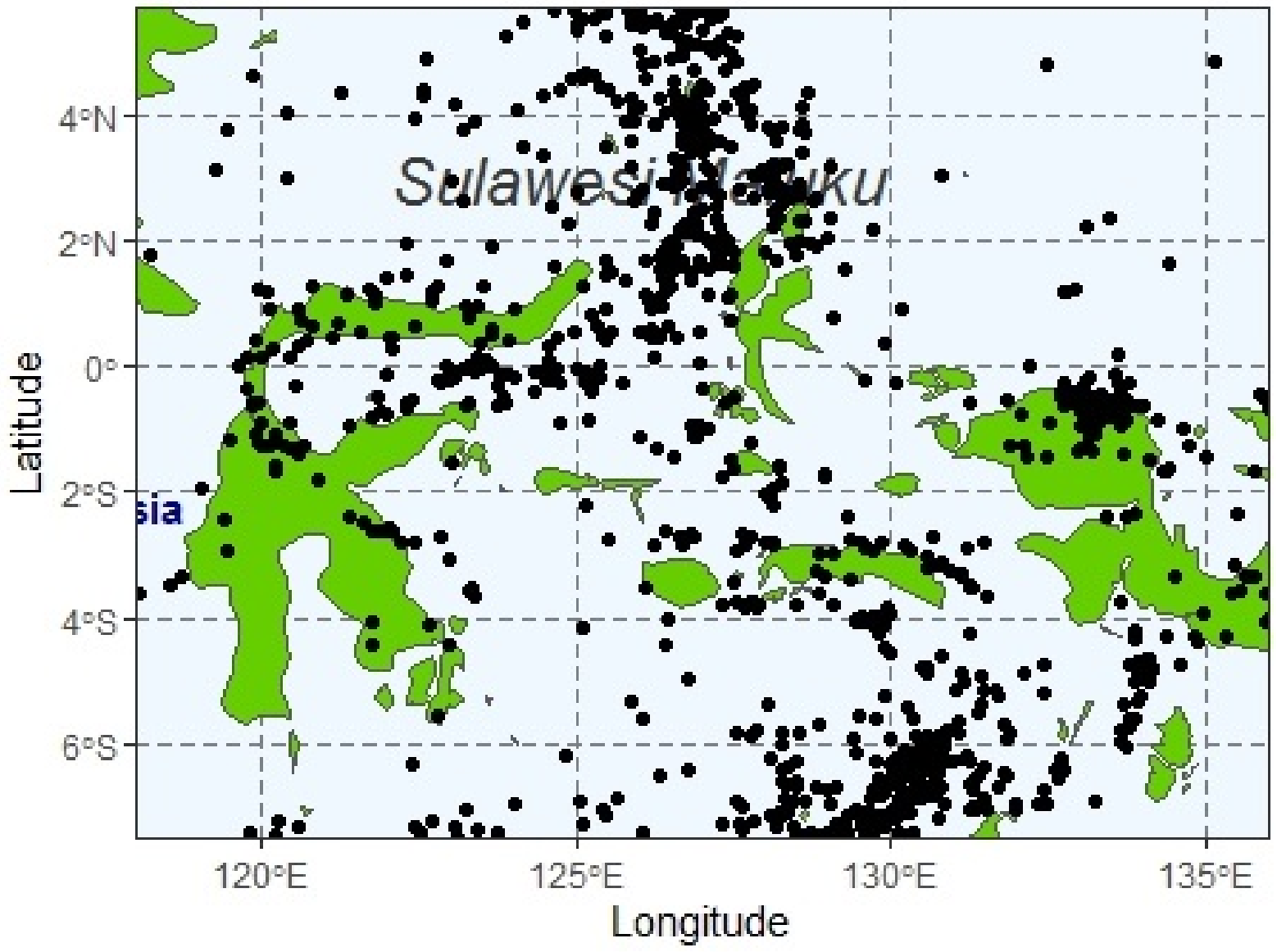}}
		\caption{The predicted intensity maps for the earthquake data in Sulawesi and Maluku by the  (a) Cauchy  and (b) variance-Gamma  cluster models, and (c) the locations of major earthquake occurrences (M $\geq$ 5) in Sulawesi and Maluku during 2009-2018 .}
		\label{pred:kppm}
	\end{figure*}
	
	The resulting estimators by the Cauchy and variance-Gamma cluster processes are depicted in Tables~\ref{table:est.result}-\ref{table:varGam2cov}. For sake of brevity, we denote in this section fault, subduction, and volcano the distances to the closest fault, subduction, and volcano. The clustering parameters discussed in this section are estimated by maximum second-order composite likelihood.
	
	To quantify the effect of geological factors, we simply assess the resulting $\bbeta$ estimates given in Table~\ref{table:varGam2cov}. We first find that major earthquake distribution in Sulawesi-Maluku is not significantly affected by a fault with significance level 0.05. The faults in Sulawesi and Maluku may not trigger major earthquakes as much as subductions and volcanoes since the majority of the great earthquakes occur at the area close to subduction zones and volcanoes in the north and south area of Sulawesi and Maluku (Figure~\ref{fig:plate} bottom). Second, the regression estimators for both subduction and volcano are negative (Table~\ref{table:varGam2cov}), which means that the closer the area to the subduction or volcano, the more likely the major earthquake to occur. In more detail, in the area with a distance 100 km closer to a subduction zone (resp. a volcano), the risk for a major earthquake occurrence increases 1.44 (resp. 1.32) times. This supports the previous studies \citep{bilek2018subduction,eggert2009volcanic}. This also indicates that an area close to the subduction tends to be around 8\% riskier for a major earthquake occurrence than the one close to the volcano.
	
	Table \ref{table:est.result} reports the estimated clustering parameters. By the Cauchy cluster point process model, the aftershocks are distributed around 84 estimated mainshocks ($\hat \kappa \times |D|, 0.28 \times 297.75$) with scale 19 km. Meanwhile the estimated mainshocks by the variance-Gamma cluster model are 120 and the aftershocks are distributed around mainshock with scale 28 km. Even though the Cauchy cluster model is slightly more clustered (smaller estimated number of mainshocks and smaller scale) than the variance-Gamma model, we do not observe a major difference (see also the envelope $K$-functions in Figures~\ref{env:kppm}c-d and the predicted intensity maps in {Figures~\ref{pred:kppm}a-b)}.
	
	{Figures~\ref{pred:kppm}(a)-(b)} present the predicted intensity maps of the Cauchy and variance-Gamma cluster models for the Sulawesi and Maluku earthquake data. The area with a very high earthquake risk is mostly predicted in two regions: (1) the northern central especially in the top tip of between Sulawesi and Maluku and (2) the southern ocean area. This is mainly because there exist major subduction zones (North Sulawesi, West and East Molucca in the north and Banda Sea subduction and Wetar Back Arc in the southern ocean) and volcanoes (Sangihe and Halmahera in north and Timor in south part) in both areas. An important note is that the predicted intensity map underestimates a number of major earthquakes at  the small area in the east (the head of Papua island, see Figure~\ref{pred:kppm}c) possibly due to the non-existence of volcano.

	\section{Concluding remarks} \label{sec:conl}
	
	In this paper, we propose to model the major earthquake (M$\geq$5) in Sulawesi and Maluku using the inhomogeneous Cox processes which involve geological variables such as the subduction zone, fault, and volcano. This study supplements the ones considering inhomogeneous Gibbs point processes \citep{anwar2012implementation,siino2017spatial}. We demonstrate that inhomogeneous cluster models improve both the independence case (Poisson model) and stationary cluster ones. We further observe that Cauchy and variance-Gamma processes fit well the earthquake distribution in Sulawesi and Maluku.
	
	Among the three geological variables considered in this study, we detect that fault is an insignificant variable. The active faults in Sulawesi and Maluku exist in the middle area. Meanwhile most of the major earthquakes occur in the north and south where major subduction zones and volcanoes exist. It could be of interest to distinguish the middle part of Sulawesi and Maluku in the analysis for further investigation. Moreover, if faults in Sulawesi and Maluku trigger earthquakes with lower magnitudes or certain depth, a study covering this phenomenon could be another research direction.
	
	
	To quantify the effect of geological variables, we focus on parametric intensity model (see Sections~\ref{sec:NSCP}-\ref{sec:LGCP}). This provides an intuitive model interpretation and ease of parameter estimation. To involve geological factors in the analysis from different perspectives, one could consider for future research the non parametric estimation \citep[e.g.][]{baddeley2012nonparametric} or semi parametric modeling such as generalized additive models \citep{youngman2017generalised}. Finally, we perform backward elimination to select the important geological variables. If more covariates are available (for example the land slope, soil density, and physical rock characteristics) and more complex multivariate dependence structure exists due to e.g. earthquake magnitudes and depths, regularization techniques \citep[e.g.][]{choiruddin2018convex, choiruddin2020regularized,choiruddin2020information} could be an option for future study.
	
	\section*{Acknowledgements}
	The research is supported by Institut Teknologi Sepuluh Nopember 848/PKS/ITS/2020.

	\bibliographystyle{plainnat}
	\bibliography{Earthquake}
	
\end{document}